\documentclass[]{aa}
\usepackage{graphicx}
\usepackage{subcaption}
\usepackage{mwe}

\usepackage{txfonts}

\usepackage{hyperref}
\usepackage{url}
\usepackage{xcolor}
\hypersetup{
colorlinks,
linkcolor={red!80!black},
citecolor={blue!80!black},
urlcolor={blue!80!black}
}

\def\cm3{cm$^{-3}$}
\def\kms{km~s$^{-1}$}

\def\msun{M$_{\odot}$}

\def\one{\ts {\,\sc i}}
\def\two{\ts {\,\sc ii}}

\def\beq{\begin{equation*}}
\def\eeq{\end{equation*}}

\def\lesssim{\mathrel{\hbox{\rlap{\hbox{\lower4pt\hbox{$\sim$}}}\hbox{$<$}}}}
\def\gtrsim{\mathrel{\hbox{\rlap{\hbox{\lower4pt\hbox{$\sim$}}}\hbox{$>$}}}}
\def\mic{\,$\mu$m}

\def\one{{\,\sc i}}
\def\two{{\,\sc ii}}

\newcommand{\code}[1]{\texttt{#1}}

\newcommand{\cmfgen}{\code{CMFGEN}}
\newcommand{\cmfflux}{\code{CMF\_FLUX}}

\def\ergs{erg\,s$^{-1}$}

\newcommand{\iso}[2]{\ensuremath{^{#1}\rm{#2}}}

\def\aj{AJ}

\def\apj{ApJ}
\def\apjs{ApJS}
\def\apjl{ApJL}
\def\aap{A\&A}
\def\araa{ARA\&A}

\def\mnras{MNRAS}
\def\nat{Nature}

\def\nifs{\iso{56}Ni}

\def\kidoub{K\one\,$\lambda\lambda$\,$7665,\,7699$}

\def\caiidoub{[Ca\two]\,$\lambda\lambda$\,$7291,\,7323$}

\def\oidoub{[O\one]\,$\lambda\lambda$\,$6300,\,6364$}

\def\mgiidoub{Mg\two\,$\lambda\lambda$\,$2795,\,2802$}

\def\nad{Na\one\,$\lambda\lambda\,5896,5890$}

\begin{document}

\title{Radiative-transfer models for dusty Type II supernovae}
\titlerunning{Radiative-transfer models for dusty Type II SNe}

\author{
Luc Dessart\inst{\ref{inst1}}
\and
D. John Hillier\inst{\ref{inst2}}
\and
Arkaprabha Sarangi\inst{\ref{inst3},\ref{inst4}}
}

\institute{
Institut d'Astrophysique de Paris, CNRS-Sorbonne Universit\'e, 98 bis boulevard Arago, F-75014 Paris, France\label{inst1}
\and
    Department of Physics and Astronomy \& Pittsburgh Particle Physics,
    Astrophysics, and Cosmology Center (PITT PACC),  \hfill \\ University of Pittsburgh,
    3941 O'Hara Street, Pittsburgh, PA 15260, USA.\label{inst2}
\and
Indian Institute of Astrophysics, 100 Feet Rd, Koramangala, Bengaluru, Karnataka 560034, India\label{inst3}
\and
DARK, Niels Bohr Institute, University of Copenhagen, Jagtvej 155A, 2200 Copenhagen, Denmark\label{inst4}
 }

   \date{Received; accepted}

  \abstract{
Dust is expected to form on a year timescale in core-collapse supernova (SN) ejecta. Its existence is revealed through an infrared brightening, an optical dimming, or a blue-red emission-line profile asymmetry. To investigate how the dust location and amount impact observations, we computed ultraviolet-to-optical spectra of interacting and standard, noninteracting Type II SNe using state-of-the-art models -- for simplicity we adopted 0.1\mic\ silicate grains. These models account for the full ejecta and treat both radioactive decay and shock power that arises from interaction of the ejecta with circumstellar material. In a Type IIn SN such as 1998S at one year, approximately $3 \times 10^{-4}$\,\msun\ of dust within the dense shell reproduces the broad, asymmetric H$\alpha$ profile. It causes an optical dimming of $\sim$\,2\,mag (which obscures any emission from the inner, metal-rich ejecta) but, paradoxically, a more modest dimming of the ultraviolet, which originates from the outer parts of the dense shell. In Type II SNe with late-time interaction, such as SN\,2017eaw, dust in the low-mass, fast outer ejecta dense shell tends to be optically thin, impacting little the optical spectrum for dust masses of order 10$^{-4}$\,\msun. In such SNe II with interaction, dust in the inner metal-rich ejecta has negligible effect on observed spectra in the ultraviolet and optical. In noninteracting SNe II, dust within the metal-rich ejecta preferentially quenches the \oidoub\ and \caiidoub\ metal lines, biasing the emission in favor of the H-rich material which generates the H$\alpha$ and Fe\two\ emission below 5500\,\AA. Our model with $5\times10^{-4}$\,\msun\ of dust below 2000\,\kms\ matches closely the optical spectrum of SN\,1987A at 714\,d. Modeling historical SNe requires treating both the ejecta material and the dust, as well as multiple power sources, although  interaction power will generally dominate.
}

   \maketitle

\section{Introduction}

There is considerable discussion about the dust cycle in the Universe -- dust  is continuously produced in stars and supernovae (SNe; \citealt{dwek_dust_07}; \citealt{gall_dust_rev_11}; \citealt{szalai_vinko_13}; \citealt{matsuura_17}; \citealt{sarangi_10jl_18}; \citealt{schneider_maiolino_24}) and destroyed by energetic processes such as shocks and ionizing radiation (\citealt{jones_nuth_11}; \citealt{micelotta_dust_18}). An important site of both formation and destruction of dust is core-collapse SNe. In SN\,1987A, \citet{lucy_dust_89} demonstrated that dust formed within the metal-rich ejecta could explain both the systematic skewness of emission line profiles (i.e., with a systematic deficit of flux received from the receding part of the ejecta) and the excess attenuation of the optical brightness observed after about 500\,d post explosion. Through modeling of the emission profile skewness, dust masses have been inferred for large SN samples (see, e.g., \citealt{niculescu_duvaz_dust_22}). Dust can also be inferred from excess infared emission as captured by the JWST in numerous historical core-collapse SNe such as SN\,1980K \citep{zsiros_80K_24}, SN\,2004et and 2017eaw  \citep{shahbandeh_jwst_23}, or SN\,2005ip (\citealt{shahbandeh_05ip_24}; see also \citealt{bevan_05ip_19}).

Theoretically dust, primarily in the form of silicates, is predicted to abruptly form within the inner, metal-rich ejecta after about 500\,d in standard Type II SNe. After several years the total dust mass is believed to be a few 0.01\,\msun\ \citep{sarangi_dust_22}, and asymptote after a few decades to a few 0.1\,\msun\ -- see inferences based on far-IR and sub-millimeter observations of SN\,1987A (e.g., \citealt{matsuura_dust_87A_11}; \citealt{indebetouw_87A_14}) and from the modeling of line-profile asymmetry of multiple core-collapse SNe (e.g., \citealt{bevan_dust_17}; \citealt{niculescu_duvaz_dust_22}). In ejecta interacting strongly with circumstellar material (CSM), dust formation is predicted to occur earlier after about a year, but in this case the location is within the compressed, dense shell formed at the interface between ejecta and CSM \citep{sarangi_cds_22}. In such interacting SNe dust would eventually be present in both the inner ejecta and the dense shell.

Late-time detection of dust in core-collapse SNe is intimately related to a powering source -- without a power source the SN is dark or too dim to allow its detection, irrespective of how much dust there may be. In SN\,1987A, radioactive decay of \nifs, $^{57}$Ni, and $^{44}$Ti have been the main power source during the first decade after the explosion (see, e.g., \citealt{larsson_87A_11}), making the SN bright enough to track dust formation within the metal-rich ejecta \citep{lucy_dust_89}. However, the luminosity of essentially all detected Type II SNe (which are at least 100 times further away from earth than SN\,1987A) years after explosion arises instead from the interaction between the ejecta and the preSN wind (the alternative power injection from a compact remnant is probably much rarer). This then raises the concern that the sample of infrared-bright, dust-producing SNe is not a representative sample of core-collapse SNe, but a subset with interaction.

A vivid signature of interaction in a Type II SN is the presence of a broad H$\alpha$ line several years after the explosion, as observed in SN\,2017eaw \citep{weil_17eaw_20}. A similar signature has been observed in a variety of Type II SNe and at various post-explosion phases, including SN\,1993J, 1998S, 2004et, or 2017eaw \citep{leonard_98S_00,matheson_93j_00a,shahbandeh_jwst_23} and all support the interaction scenario (see, e.g., \citealt{fransson_93j_96,fransson_cno_93J_98S_05,dessart_late_23}). The powering by interaction is instrumental for the detection of these SNe by the JWST (see, e.g.,  \citealt{shahbandeh_jwst_23}). A challenge is to discern how the dust properties inferred from such SNe undergoing interaction can be generalized to core-collapse SNe, and whether the dust forms within the metal-rich inner ejecta, the outer dense shell, or both.

In this exploratory paper, we investigate the impact of dust mass and location on the escaping radiation from Type II SN ejecta. We consider a variety of configurations that reflects the diversity of Type II SNe, thus including Type II SN ejecta strongly interacting with CSM (i.e., bona-fide interacting SNe such as SN\,1998S), ejecta weakly interacting with CSM (e.g., a SNe IIP with late-time interaction with the progenitor wind, such as SN\,2017eaw), as well as noninteracting SN ejecta in which radioactive decay is the only power source (of which the best example is SN\,1987A).

In contrast to previous work that focused primarily on the modeling of individual spectral lines (e.g., \citealt{lucy_dust_89}; \citealt{bevan_barlow_16}), we performed globally consistent radiative-transfer calculations that consider the entire ejecta (starting from a physically consistent model of the progenitor and its explosion; see, for example, \citealt{dessart_sn2p_21}) as well as the various power sources influencing its energy content. Where appropriate, we included a treatment of shock power to mimic the interaction with CSM. Our models cover from the ultraviolet to the infrared and thus address the global impact of dust across the electromagnetic spectrum as well as capture the differential effect it  has on emission from within the metal-rich inner ejecta regions or from the outer, fast-moving dense shell.

In the next section, we present the Type II SN models used as a basis for the dust radiative-transfer calculations. We then briefly summarize in Section~\ref{sect_dust} the scattering and absorptive properties of dust. In Section~\ref{sect_cmfgen}, we discuss the treatment of dust in the radiative transfer code \cmfflux\ \citep{HM98,HD12}, including benchmarking tests obtained with a Monte Carlo approach. The following three sections present the results for the three Type II SN ejecta configurations we selected, starting with an interacting SN (Section~\ref{sect_sniin}), a Type IIP SN with late-time interaction (Section~\ref{sect_sniipcsm}), and a noninteracting Type II SN (Section~\ref{sect_snii}). We present our conclusions in Section~\ref{sect_conc}. In the appendices, we provide additional discussion on the numerical treatment of dust in \cmfflux\ and benchmarking tests, as well as additional information on the ejecta properties characterizing the SN\,IIn model used in Section~\ref{sect_sniin}.

\begin{table*}
  \caption{Summary of ejecta properties used as initial conditions for our dust calculations. (See Section~\ref{sect_setup} for discussion.)
\label{tab_cmfgen_init}
}
\begin{center}
\begin{tabular}{
lccccccccc
}
\hline
Model          &   Age     & $M_{\rm ej}$  &$E_{\rm kin}$ & $M$(O) & $M$(\nifs) & $L_{\rm decay,abs}$ &   $L_{\rm sh,abs}$ & $V_{\rm CDS}$ & $M_{\rm CDS}$ \\
\hline
               &    [d]    &       [\msun] &      [erg] & [\msun] & [\msun] & [\ergs] & [\ergs]  & [\kms] & [\msun] \\
\hline
SN\,IIn        &    300    &   12.00 &   1.5(51) &  0.75 &  0.032  & 3.1(40)   & 2.0(41)  & 5000    & 2.4   \\
SN\,IIP/CSM    &    1000   &   11.49 &   9.2(50) &  1.02 &  0.063  & 3.2(37)   & 1.0(40)  & 8000    & 0.13  \\
SN\,II         &     700   &   11.36 &   9.2(50) &  1.02 &  0.063  & 8.3(38)   & \dots    & \dots   & \dots \\
\hline
\end{tabular}
\end{center}
Notes: The columns give the age, the total mass, the kinetic energy, the O and \nifs\ mass for each ejecta model. We then give the absorbed decay power in each model, and for the SN\,IIn and SN\,IIP/CSM, we also give the total shock power absorbed in the outer ejecta dense shell, whose velocity and mass are given in the last two columns. In the SN\,II model, the only power source is radioactive decay. Numbers in parentheses refer to powers of ten.
\label{table_init}
\end{table*}

\section{Model selection and properties}
\label{sect_setup}

In this study, we considered the main type II SN configurations encountered in nature, namely ejecta expanding in a vacuum and powered by radioactive decay and ejecta that interact, either weakly or strongly, with CSM. Our first model is for an interacting SN and is named ``SN\,IIn''. It originally derives from the m15 model of \citet{dessart_csm_22} but was modified in two ways. To reflect a strong interaction with CSM and match the emission line width observed in SN\,1998S at nebular times \citep{leonard_98S_00}, all regions beyond 5000\,\kms\ were piled up into a dense shell. The corresponding cumulative mass of these ejecta regions is 1.2\,\msun.  We then doubled that mass to account for the CSM mass at the origin of the ejecta deceleration and hence placed a dense shell of 2.4\,\msun\ at 5000\,\kms. We assumed no material beyond the dense shell although in practice there is some CSM but with a density orders of magnitude smaller than in the dense shell. The second adjustment was to deposit a shock power of $2\times10^{41}$\,\ergs\ within this dense shell. Although this model is analogous to those shown in \citet{dessart_csm_22}, it was specifically calculated for the present study (but with the same approach as described in that earlier work) to accommodate the stronger deceleration of the outer ejecta by a more massive CSM than adopted in that former study. The SN\,IIn model has an age of 300\,d and will be confronted to the observations of SN\,1998S at 375\,d in Section~\ref{sect_sniin}.

The second model is for a standard Type II SN interacting with its progenitor red-supergiant wind at 1000\,d after explosion. This model was taken directly from \citet{dessart_late_23} and corresponds to the s15p2 model of \citet{sukhbold_ccsn_16} and  \citet{dessart_sn2p_21} but with an  interaction power of 10$^{40}$\,\ergs\ deposited  within a low-mass (i.e., 0.13\,\msun) high-velocity (i.e., 8000\,\kms) dense shell (hereafter called ``SN\,IIP/CSM'').\footnote{Here, we are not defining a new class of Type II SNe. We merely attempt to characterize an event like SN\,2017eaw, which was a standard Type IIP SN for about two years \citep{vandyk_17eaw_19} and subsequently morphed into an interacting SNe \citep{weil_17eaw_20,dessart_late_23}. As discussed by \citet{dessart_late_23} and as far as the physical process of interaction is concerned, the IIn SN type is far too restrictive and captures only a small fraction of H-rich ejecta interacting with CSM. Naming this model a IIP/IIn is inadequate because at late times SN\,2017eaw exhibits broad rather than narrow lines.} In \citet{dessart_late_23}, this model was just named Pwr1e40 since all their simulations were based on the same progenitor and explosion model s15p2. Our present SN\,IIP/CSM model at 1000\,d will be confronted to the observations of SN\,2017eaw at 900\,d in Section~\ref{sect_sniipcsm}.

\begin{figure}
\centering
\includegraphics[width=0.49\textwidth]{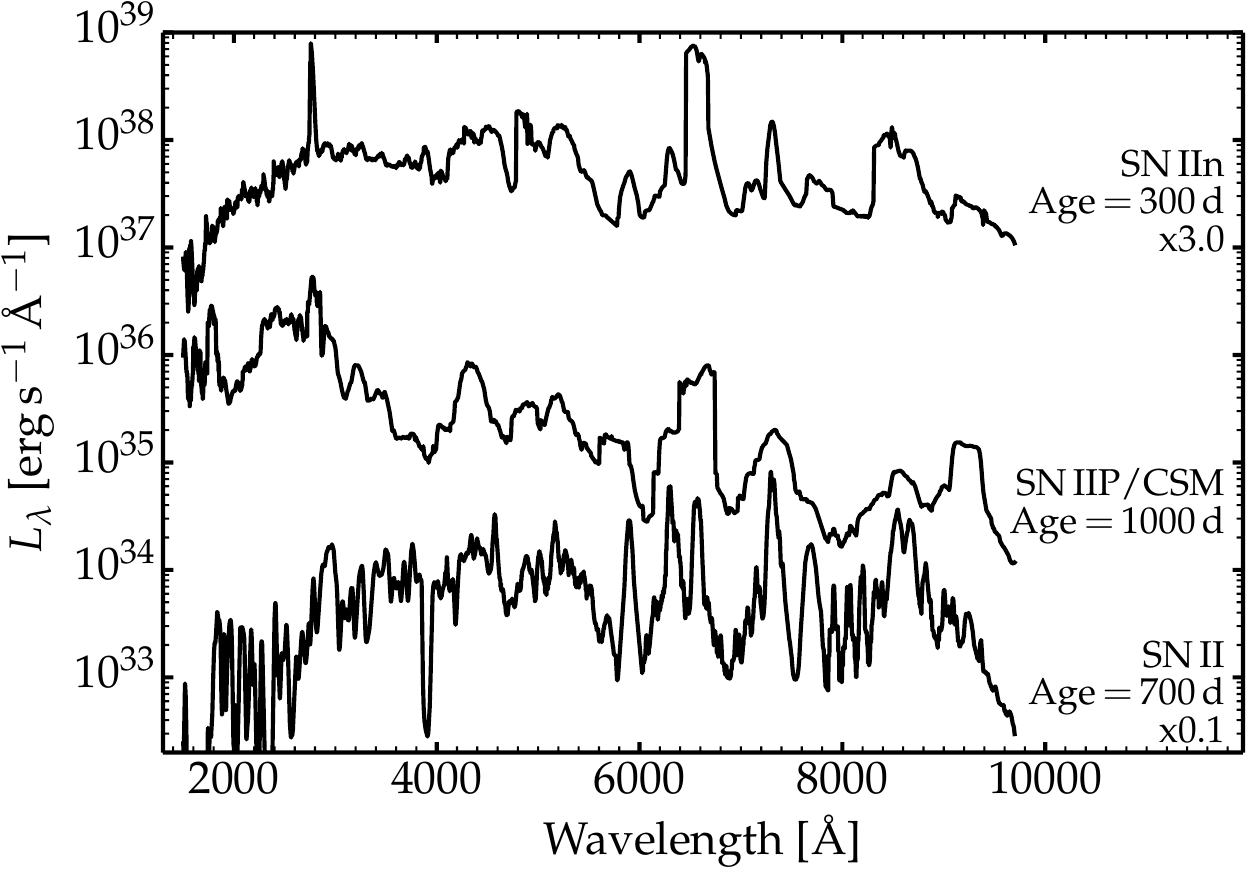}
\caption{Luminosity versus wavelength in the absence of dust for our models SN\,IIn (top), SN\,IIP/CSM (middle), and SN\,II (bottom). Some fluxes have been scaled for better visibility (see label ``x''). Ages span 300 to 1000\,d after explosion. [See also Table~\ref{table_init} and Section~\ref{sect_setup}.]
\label{fig_set}
}
\end{figure}

The third and last model is for a noninteracting and thus standard type II SN. It corresponds to the unadulterated model s15p2 of \citet{sukhbold_ccsn_16} and  \citet{dessart_sn2p_21} and it was computed at an age of 700\,d. The method of calculation is the same as in \citet{dessart_sn2p_21} except for the different age of 700\,d. This model is named SN\,II and will be confronted to the observations of SN\,1987A at 714\,d in Section~\ref{sect_snii}. Arguably, SN\,1987A was a peculiar Type II SN and derived from a blue-supergiant rather than a red-supergiant progenitor but this has little relevance at nebular times. One discrepancy with SN\,1987A was the solar metallicity adopted in model s15p2 whereas a metallicity of one third to one half solar would be needed to accommodate this SN in the Large Magellanic Cloud.

Our set is thus composed of models that we name SN\,IIn, SN\,IIP/CSM, and SN\,II, whose properties are summarized in Table~\ref{table_init}. These ejecta are characterized by similar inner-ejecta properties in terms of density and composition but differ in the power sources. Radioactive decay is included in all three ejecta where its influence is limited to the inner, metal-rich regions. It is the only power source in model SN\,II. An additional shock power is deposited in the outer ejecta of models SN\,IIn and SN\,IIP/CSM with a magnitude of $2\times10^{41}$\,\ergs\ and 10$^{40}$\,\ergs\ -- these interaction powers are ten to several hundred times greater than the contemporaneous radioactive-decay power.

We show the corresponding spectra for these three models (in the absence of dust) in Fig.~\ref{fig_set} -- additional information is provided in Fig.~\ref{fig_m15_300d_prop} for the model SN\,IIn and in \citet{dessart_csm_22} and \citet{dessart_late_23} for the other two models. The SN\,IIn spectrum is the most luminous, with the bulk of the emission arising from the outer dense shell at 5000\,\kms. Emission occurs primarily through lines rather than continuum processes and in the absence of dust the model flux falls primarily in the optical. The older more weakly interacting model SN\,IIP/CSM has more flux in the ultraviolet (with strong Ly$\alpha$, not shown) and most of the emission comes from lines forming in and around the outer ejecta dense shell at 8000\,\kms. The third, interaction-free model exhibits totally different properties since most of the emission arises from the inner, metal-rich ejecta and is dominated by optical forbidden lines -- this emission from the decay-powered inner ejecta is also present in the other two models but swamped (though not masked) by the contribution associated with the interaction.

Other model choices, parameters and power sources are possible. However, our goal is only to capture some of the variations that may be found in Type II SN ejecta at late times. With these three models, we capture the essence of the three associated categories, namely, interacting SNe, standard SNe with weak late-time interaction, and noninteracting SNe in which the only source of power is radioactive decay -- that latter configuration may only have one observed instance at such late times with SN\,1987A due to its proximity and the fast, tenuous wind of its blue supergiant progenitor. After describing in the next two sections the dust opacities and the treatment of dust absorption, scattering, and emission in \cmfflux, we discuss the impact of dust on the escaping radiation associated with these three SN ejecta configurations, in particular in relation to the dust location and abundance.

\begin{figure}
\centering
\includegraphics[width=0.49\textwidth]{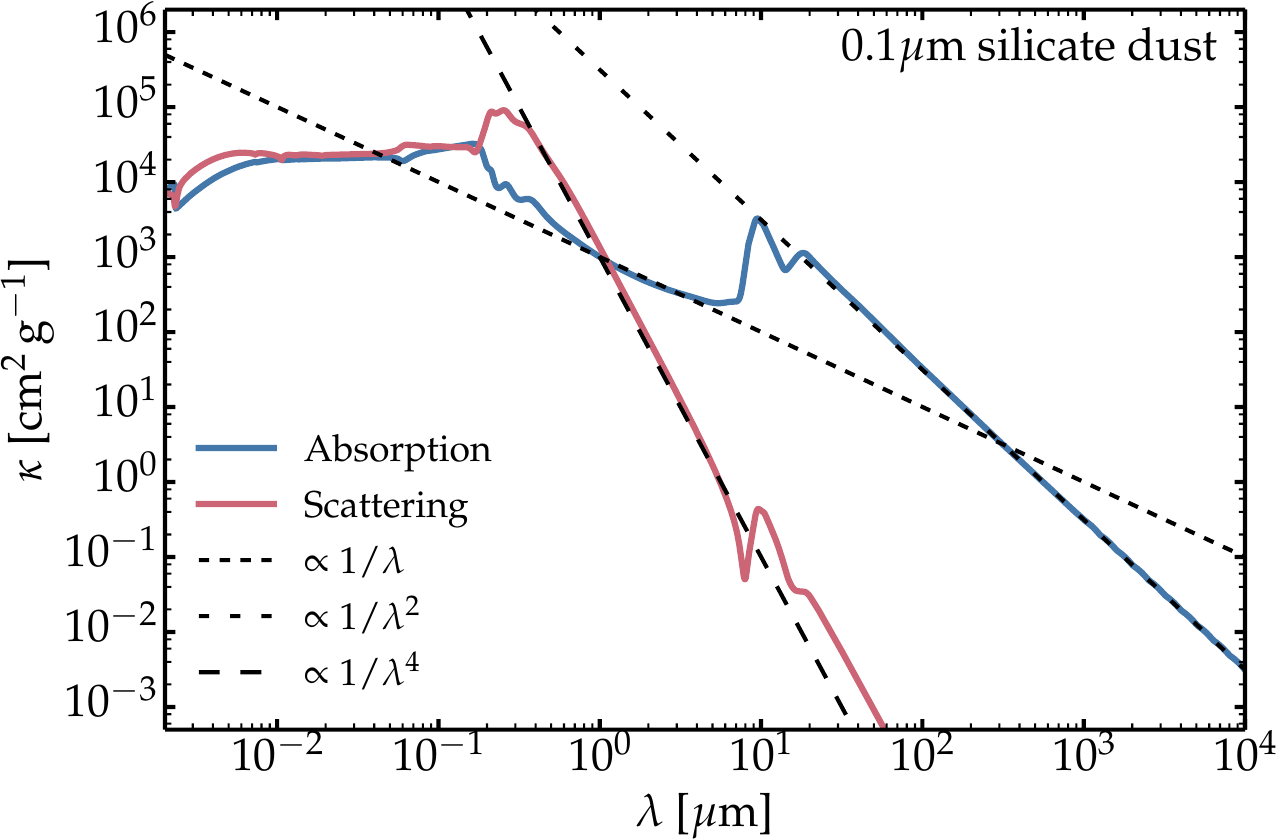}
\caption{Illustration of the wavelength dependence of the absorptive (blue) and scattering (red) opacity of the 0.1\,$\mu$m silicate dust grains adopted in all radiative-transfer calculations with dust presented in this work. Dashed lines indicate the slopes in various spectral regions. [see Section~\ref{sect_dust} for discussion.]
\label{fig_dust_chi_sil}
}
\end{figure}

\section{Dust opacities}
\label{sect_dust}

The dust opacities used in this work are taken from previous work. For simplicity, and also because silicates are the first dust grains to form in SN ejecta \citep{sarangi_dust_22}, we consider only one type of dust grains --- 0.1\mic\ silicates --- and take silicate opacities from \citet{draine_li_07} and \citet{weingartner_draine_01}. Tests indicate that other choices yield similar qualitative results (thus not shown) and are  not critical for what concerns us here. They will become relevant when modeling specific SNe in detail and when comparing models with high-quality observations that cover ultraviolet to infrared wavelengths.

Figure~\ref{fig_dust_chi_sil} shows the absorptive and scattering opacity of 0.1\mic\ silicate grains versus wavelength. Dust opacities in the optical are typically a thousand times greater than that due to electron scattering and one may thus expect the influence of dust opacity in nebular-phase SN ejecta, even when the electron-scattering optical depth is below unity. One may also expect strong implications in the optical even for modest dust masses.

For wavelengths below the grain size, the opacity is essentially constant and the scattering and absorptive opacities are comparable. In the optical the scattering opacity can substantially exceed the absorptive opacities at some wavelengths. For wavelengths about ten times larger than the grain size, the opacity drops as $1/\lambda^\beta$, with $\beta$ of a few. The decrease in opacity with wavelength is greater for the scattering opacity, which goes typically as $1/\lambda^4$ and thus similar to the dependence obtained for Rayleigh scattering of optical radiation by molecules in the earth atmosphere \citep{rayleigh_1899}. Absorptive dust opacities decline less steeply with wavelength and scale roughly as $1/\lambda^2$ (the curve is not a straight line but has distinctive bumps). In the mid-infrared regions, the opacities can still be large (especially for larger grains) and comparable to those found in the optical so dust emission in the infrared may be affected by optical depth.

\begin{figure}[h]
\centering
\includegraphics[width=0.49\textwidth]{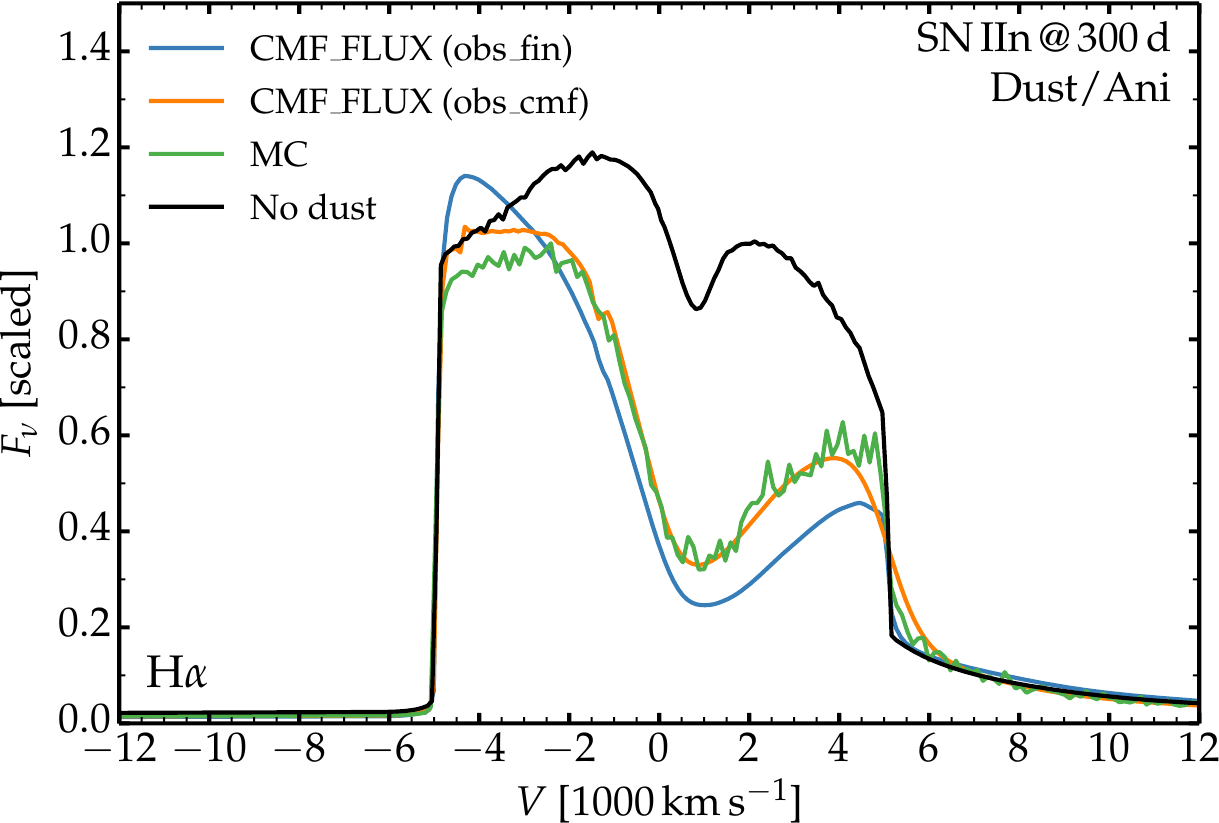}
\caption{Comparison of the SN\,IIn model H$\alpha$ profile between the MC and \cmfflux\ calculations when we assume anisotropic scattering ($g=$\,0.8). From \cmfflux, we show both the CMF (i.e., obs\_cmf) and the observer's frame (i.e., obs\_fin) calculations. The no-dust case, computed in the observer's frame, is shown in black. In all cases shown here, only H$\alpha$ is included in the profile calculation (i.e., there is no background flux arising, for example, from the overlap with a forest of lines from Fe\one\ or Fe\two).  
\label{fig_comp_mc_cmfgen_gp8}
}
\end{figure}

\begin{figure}[h]
\centering
 \includegraphics[width=0.49\textwidth]{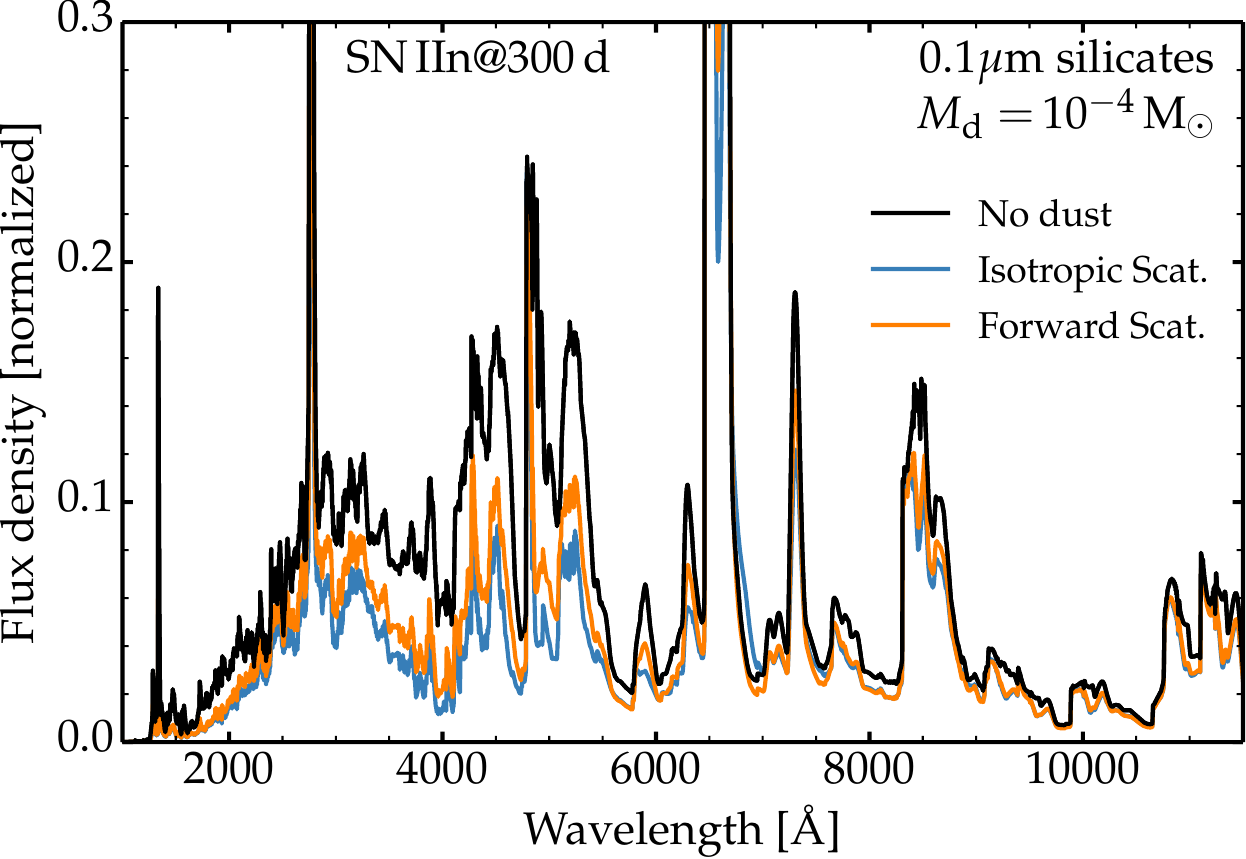}
\caption{Comparison of isotropic versus forward (i.e., $g=$\,0.8) dust scattering for the SN\,IIn model discussed in Section~\ref{sect_sniin}. For all three cases, we show the results of the CMF calculation with \cmfflux\ in which anisotropic dust scattering is properly taken into account. [See Section~\ref{sect_cmfgen} and Appendix\,\ref{append_HG} for discussion.]
\label{fig_gp0_vs_gp8}
}
\end{figure}

\section{Treatment of dust in \cmfflux}
\label{sect_cmfgen}

In this preliminary study we only allow for the influence of dust on the computed spectrum. We accurately treat the  wavelength-dependent absorption and scattering cross-sections of the dust but, as noted earlier, we consider only a single grain size and type. For wavelengths smaller than or comparable to the grain size the scattering is anisotropic. Dust scattering is generally forward-peaked, and treating it and its full wavelength dependence would be difficult. We therefore assume two limiting approximations -- isotropic scattering, and forward scattering assuming a Henyey-Greenstein phase function (Appendix~\ref{append_HG}). To test the accuracy of the calculations we use two approaches -- a formal solution of the radiative transfer equation (as implemented in \cmfflux; \citealt{HM98}; \citealt{HD12}) and a Monte-Carlo (MC) calculation \citep{hillier_91,DH11_pol}.

The implementation of isotropic scattering into the \cmfflux\ profile calculations is simple. To compute the emissivity we assume that the dust scattering is coherent in the comoving frame, which is an excellent approximation since the random motions of the dust particles will be small ($<$\,1\,\kms). For isotropic scattering the dust scattering emissivity can be solved for using the moment equations or by using a lambda iteration.

The implementation of anisotropic dust scattering into the MC code is straightforward, but is much more difficult for the formal solution. Its implementation is described in Appendix~\ref{append_HG}. Due to the complex angle and wavelength dependence of the dust emissivity, we only compute the observed spectrum via a comoving-frame (CMF; the resulting spectrum is named ``obs\_cmf'') approach. That is, we compute  the intensity at the outer boundary of our model in the CMF, and then transform it into the observer's frame. Generally we also compute the observed spectrum using an observer's frame calculation using the mean intensities and hence emissivities that were computed in the comoving frame (the resulting spectrum is named ``obs\_fin''). This later approach works well for isotropic scattering, but does not work for an arbitrary phase function (since the scattered intensities cannot be treated as a function of the mean intensity). The agreement between the two approaches is generally excellent although the  observed spectrum computed using the CMF approach typically shows a slight bleeding of line profiles to longer wavelengths due to numerical diffusion (although line equivalent widths are well conserved). The wavelength bleeding arises because in the CMF calculation the radiation has to be transported in both depth and wavelength space.

The dust temperature is currently assumed rather than calculated. For simplicity, we adopt a single temperature for the dust. This limits the consistency of the emission in the infrared, which is not the main focus here. In the future, dust will need to be incorporated within the \cmfgen\ calculation and fully coupled to the gas and radiation through absorption, scattering, and emission (collisional heating is negligible; \citealt{sarangi_dust_22}).

For the calculations presented in this work, we consider three different ejecta models (see Section~\ref{sect_setup}) and associated \cmfgen\ results for the level populations, electron density, and temperature. We then compute the spectrum by adding dust with specifications for its spatial distribution within the ejecta, the total dust mass, the type of dust (e.g., silicates) and grain size. For the dust distribution we currently adopt a gaussian profile with a parametrized center and width, or we assume the dust is uniformly distributed within a volume bounded between the ejecta base and some specified ejecta velocity (e.g., 2000\,\kms, which is the rough location of the outer edge of the metal-rich layers in Type II SN ejecta).

In Fig.~\ref{fig_comp_mc_cmfgen_gp8} we show a spectrum comparison for a case of highly anisotropic scattering for the SN\,IIn model at 300\,d and in which the dust is introduced in the dense shell (see Appendix~\ref{append_dc} and Section~\ref{sect_sniin}).  The profiles obtained with the MC code and the CMF calculation with \cmfflux\ (i.e., obs\_cmf)  are in good agreement -- especially the wings. With anisotropic scattering, there is less suppression of the red peak, although the basic profile shapes are similar. As expected the observer's frame calculation from \cmfflux\ (i.e., obs\_fin) differs since in that frame the dust scattering is still treated isotropically. For practical calculations of line profiles isotropic scattering will generally yield adequate results, especially considering the uncertainties in dust properties, the location of the dust in the ejecta, and the amount of dust. In the present case the use of isotropic scattering would simply lead to a slight underestimate of the dust mass (i.e., a greater attenuation results for a given dust mass).

Indeed, tests show that assuming isotropic or anisotropic scattering has only a mild impact on the resulting SN spectrum (see Fig.~\ref{fig_gp0_vs_gp8}). Scattering opacities are low in the infrared so the scattering anisotropy is primarily important for the optical and ultraviolet. When the dust is optically thin, dust has a negligible role on the SN spectra (i.e., line profiles) and photometry so the dust phase function is irrelevant. Conversely, when the dust is optically thick, photons scatter multiple times, which tends on average to make their propagation more isotropic. Consequently, most calculations in this work assume isotropic scattering. We  always account for both absorptive and scattering opacity. In Appendix~\ref{append_dc} we  provide additional test calculations that illustrate both the accuracy of the profile calculations and the influence of anisotropic scattering. In most cases, we confront the results from the MC and from the \cmfflux\ calculations.

\section{Dust in the SN\,IIn model and comparison to SN\,1998S at 375\,d}
\label{sect_sniin}

In this section, we consider various choices for the dust location and mass in the ejecta of the SN\,IIn model and discuss their impact on the spectral and photometric properties. We then confront a selection of such results to the observations of SN\,1998S at 375\,d.

\begin{figure}
    \centering
    \begin{subfigure}[b]{0.4\textwidth}
       \centering
       \includegraphics[width=\textwidth]{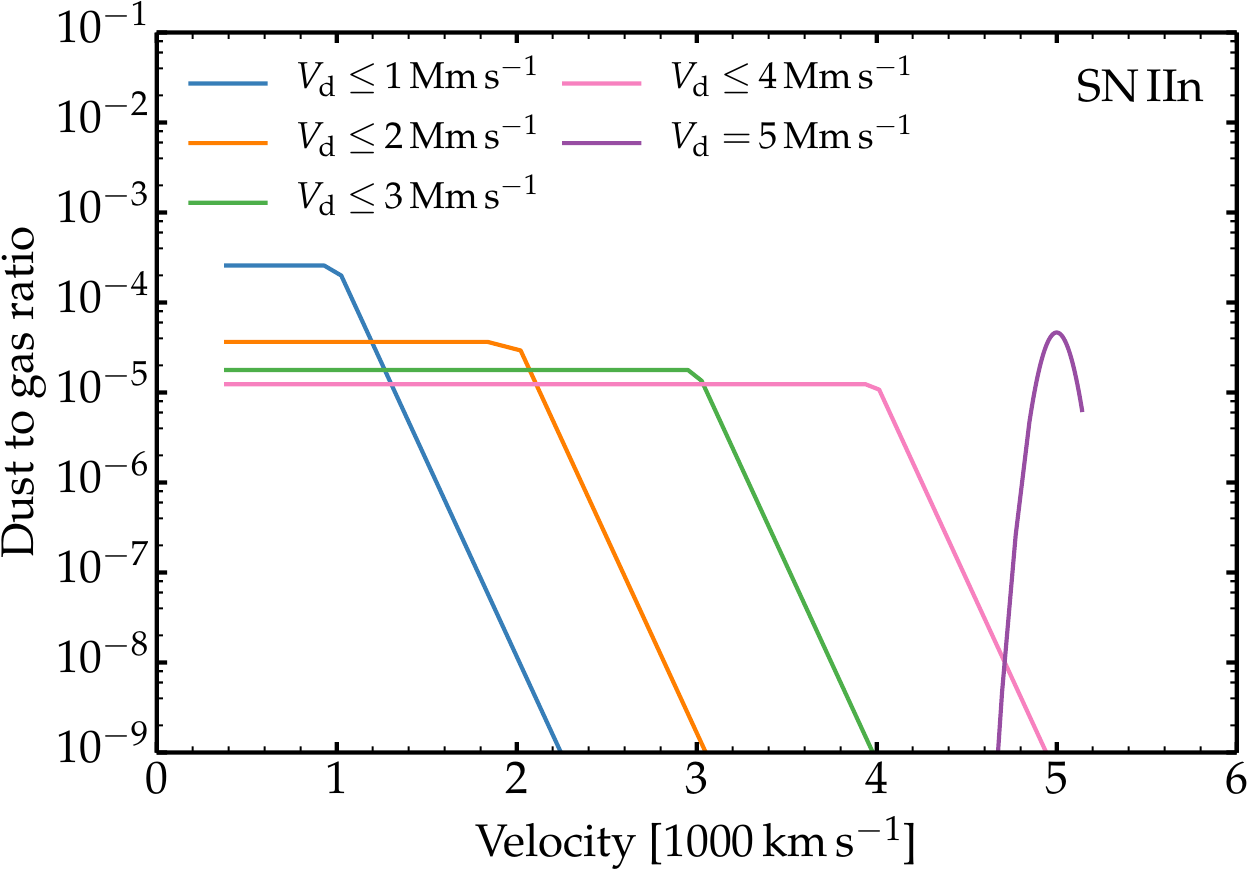}
    \end{subfigure}
    \hfill
    \centering
    \begin{subfigure}[b]{0.4\textwidth}
       \centering
       \includegraphics[width=\textwidth]{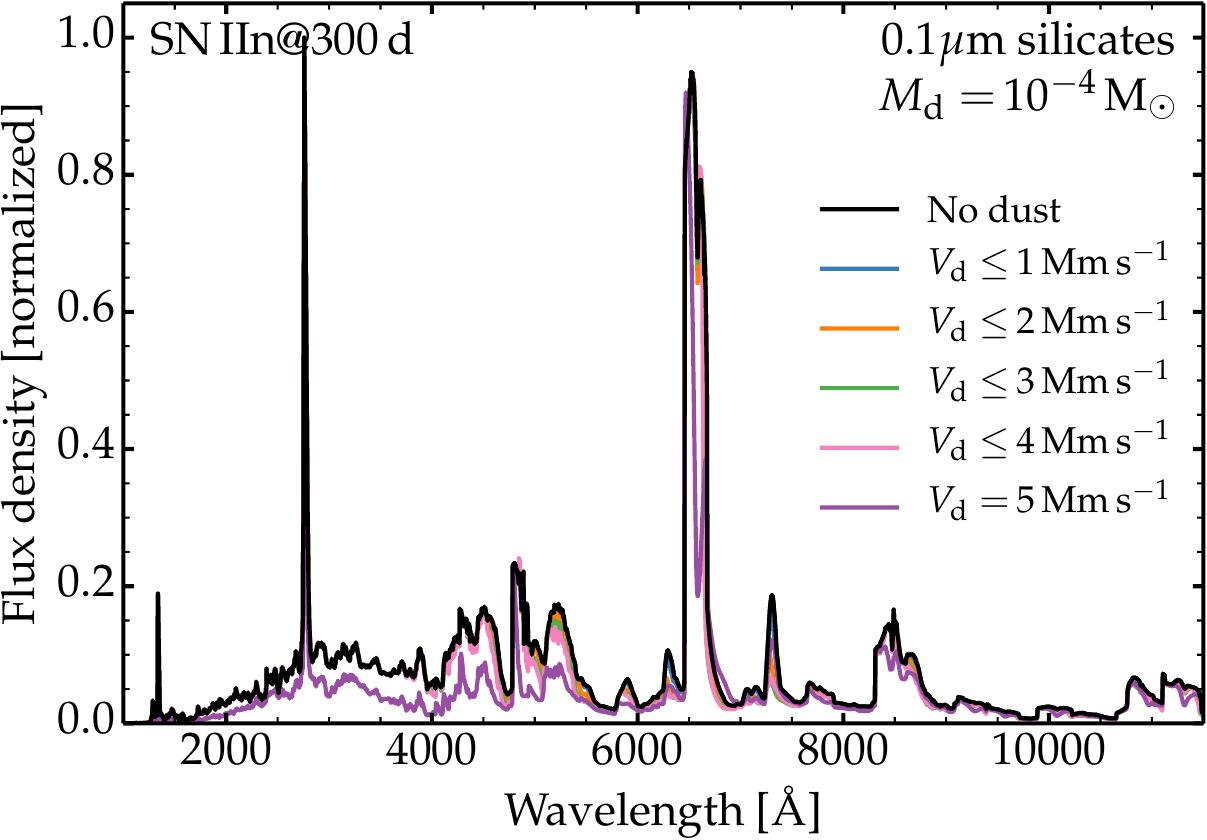}
    \end{subfigure}
    \hfill
    \centering
    \begin{subfigure}[b]{0.4\textwidth}
       \centering
       \includegraphics[width=\textwidth]{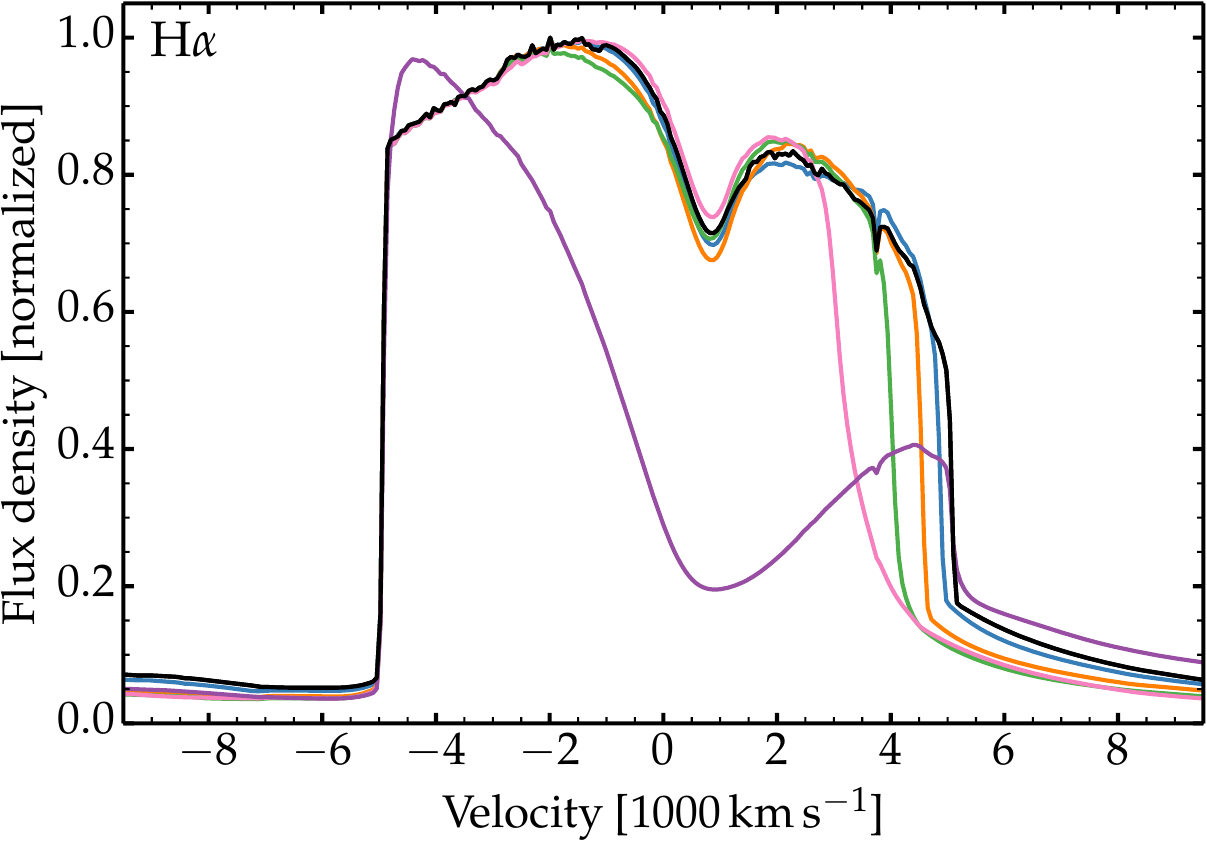}
    \end{subfigure}
\caption{Impact of the dust location on the spectral properties of the SN\,IIn model at 300\,d. We show various choices for the dust spatial distribution (top) and the corresponding predictions for the resulting optical spectrum (middle) and the H$\alpha$ profile (bottom). The dust is made of 0.1$\mu$m silicate grains and the total dust mass is 10$^{-4}$\,\msun. [See Section~\ref{sect_sniin} for discussion.]
\label{fig_comp_loc_sil_sniin}
}
\end{figure}

Figure~\ref{fig_comp_loc_sil_sniin} illustrates the impact on the ultraviolet and optical spectra (middle panel) and on H$\alpha$ (bottom panel) for various dust distributions (top panel) -- the total dust mass adopted is fixed here to 10$^{-4}$\,\msun\ and we use 0.1\mic\ silicates. The presence of dust in the inner ejecta has only a modest impact on the spectrum. If bounded within 1000\,\kms, the influence is negligible. If the outer edge of that dust-rich region is located further out at 2000 or 3000\,\kms, the strength of metal lines like \oidoub\ is reduced but these lines are typically quite weak in this model since the bulk of the emission arises from the outer ejecta dense shell at 5000\,\kms\ (the amount of decay power absorbed in the inner ejecta is a tenth of the shock power absorbed; see Table~\ref{table_init}).

Since the bulk of the H$\alpha$ emission originates from the outer shell, the influence of the dust on H$\alpha$ is small when the outer edge of the dust-rich region is confined to small velocities (e.g., 1000\,\kms). However, when the edge of the dust-rich region is shifted to 3000\,\kms\ a flux deficit on the red side of H$\alpha$ is apparent, and becomes large as we shift that outer edge to even higher velocities. This arises from the depletion of H$\alpha$ photons emitted by the rear shell. As  the dust is interior to the shell the blue side of the profile is ``unaffected'' as are photons emitted from the rear shell that do not pass through the dust.

A drastic and strongly wavelength dependent alteration to the spectrum is obtained when the dust is placed within the dense shell at 5000\,\kms. The impact in the ultraviolet is modest because the ultraviolet radiation, being already strongly attenuated by the mere presence of gas in the dense shell, tends to form in the outer parts of the dense shell and hence is not strongly attenuated by the dust. \mgiidoub\ already exhibits a strong blue-red asymmetry without dust and adding in dust makes little difference. In the optical, the flux is strongly attenuated at shorter wavelengths due to the larger dust opacity of 0.1\mic\ silicate grains in this spectral range (see Fig.~\ref{fig_dust_chi_sil}) --- the continuum at longer wavelengths (e.g., 8000\,\AA) is only weakly affected. However, the impact on the H$\alpha$ profile is large as evidenced by the strong flux deficit on the red side. As for the dust-free case, there is a greater attenuation around line center because of the greater pathlength of photons emitted from the limbs of the spherical dense shell and thus having a near-zero projected velocity.

\begin{figure}
   \centering
    \begin{subfigure}[b]{0.45\textwidth}
       \centering
\includegraphics[width=\textwidth]{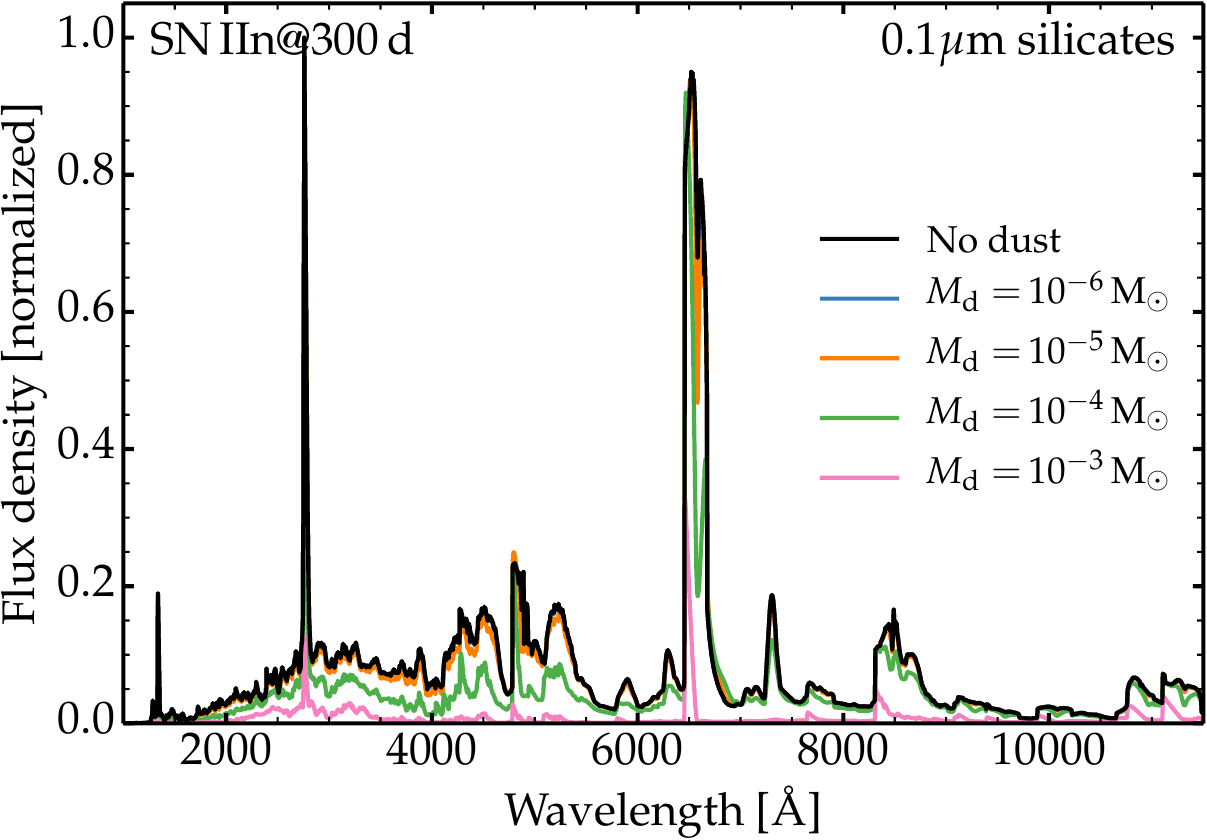}
    \end{subfigure}
    \hfill
    \centering
    \begin{subfigure}[b]{0.45\textwidth}
       \centering
       \includegraphics[width=\textwidth]{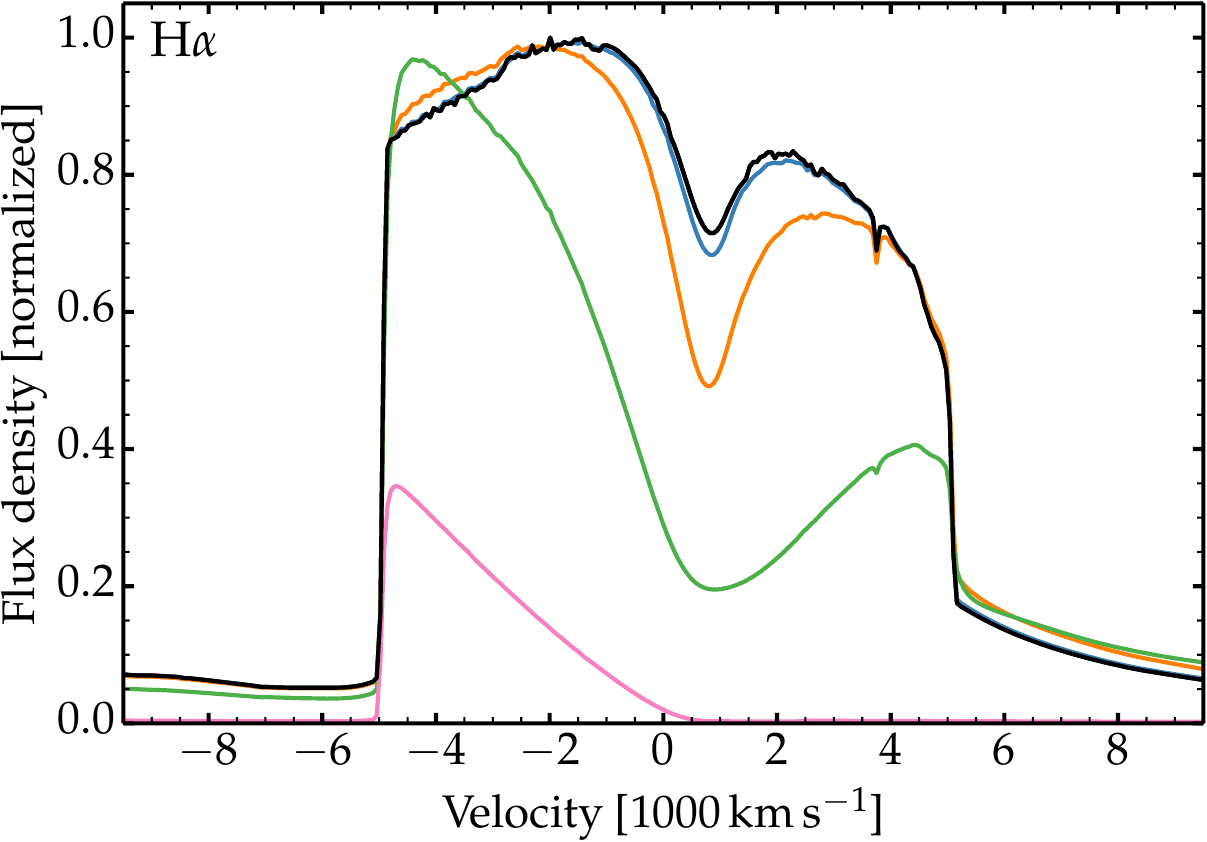}
    \end{subfigure}
\caption{Impact of the dust mass on the spectral properties of the SN\,IIn model at 300\,d. When included, the dust is placed within the dense shell at 5000\,\kms\ (purple curve in the top panel of Fig.~\ref{fig_comp_loc_sil_sniin}). The figure is analogous to Fig.~\ref{fig_comp_loc_sil_sniin} and shows the impact on the ultraviolet and optical spectrum (top) and on the H$\alpha$ profile (bottom).}
\label{fig_comp_mass_sil_sniin}
\end{figure}

All the line profiles shown in the bottom panel of Fig.~\ref{fig_comp_loc_sil_sniin} exhibit an extended red wing. In the absence of dust, this excess flux at Doppler velocities greater than $V_{\rm CDS}$ (of 5000\,\kms\ here) redward of the rest-wavelength of H$\alpha$ arises from the scattering of H$\alpha$ photons emitted within the dense shell with free electrons in the dense shell. When dust is introduced in the inner ejecta, it preferentially obscures the regions that contribute line photons endowed with the greatest redshift (i.e., those coming from the back side of the dense shell) and consequently both the line flux on the red side and the extended redwing flux decrease (i.e., the scattered flux scales with the line flux). However, when dust is introduced within the dense shell, the extended redwing is then caused primarily by dust scattering. Because the dust scattering optical depth across the dense shell is about one at the H$\alpha$ wavelength (and more than ten times greater than that from electron scattering), the excess flux in the redwing is even greater than in the dust-free case (and in spite of the reduced red peak flux). This effect may be overestimated because of our assumption of isotropic dust scattering.

Figure~\ref{fig_comp_mass_sil_sniin} is a counterpart of Fig.~\ref{fig_comp_loc_sil_sniin} but showing the impact of varying the dust mass, assuming that the dust resides in the dense shell at 5000\,\kms (see the precise distribution of the dust-to-gas-ratio in the top panel of Fig.~\ref{fig_comp_loc_sil_sniin}). A mass of at least 10$^{-5}$\,\msun\ is needed to sizably affect  the spectrum and H$\alpha$, while for dust masses greater than 10$^{-4}$\,\msun\ the flux deficit on the red side of H$\alpha$ is very pronounced. When the dust mass is increased to 10$^{-3}$\,\msun, the H$\alpha$ profile exhibits only a single peak, which is also blue shifted by approximately 4000\,\kms\ -- the red component is absent. For that higher dust mass, the maximum dust-to-gas ratio in the dense shell is about $5\times10^{-4}$. In terms of dust optical depth in the H$\alpha$ spectral range, it increases roughly from 0.004 to 4.0 (absorption part) and from 0.024 to 24.0 (scattering part) as we increase the dust mass from 10$^{-6}$ to 10$^{-3}$\,\msun. The transition from optically thin to optically thick conditions in the H$\alpha$ region occurs at a dust mass of $\sim$\,10$^{-4}$\,\msun\ (and at lower dust masses when considering scattering alone; see also Fig.~\ref{fig_dust_chi_sil}).

\begin{figure}
\centering
 \includegraphics[width=0.49\textwidth]{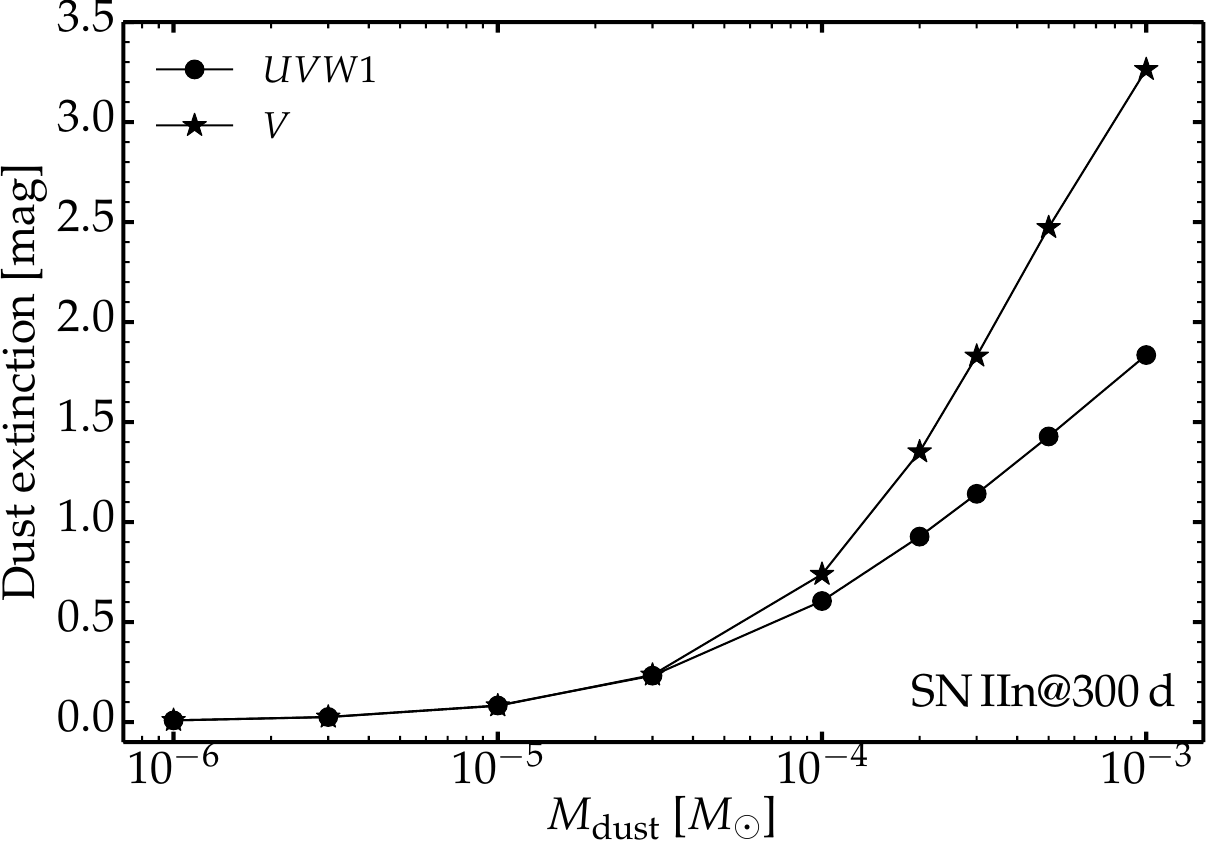}
\caption{Extinction due to ejecta dust for the SN\,IIn model. We consider dust masses between 10$^{-6}$ and 10$^{-3}$\,\msun\ and show the results for the $UVW1$-band (ultraviolet range) and $V$-band filters. As in Fig.~\ref{fig_comp_mass_sil_sniin}, the dust is placed in the outer-ejecta dense shell. [see Section~\ref{sect_sniin} for discussion.]
\label{fig_sniin_extinction}
}
\end{figure}

While a blue-red profile asymmetry is often used to assess the amount of dust present in the region of formation of certain lines, another important aspect is the magnitude of extinction associated with the presence of dust. This is important to accurately constrain the energy budget, in particular in the context of interacting SNe since it connects to the properties of the ejecta, the CSM, and the shock power. This may in principle be estimated by means of panchromatic observations covering from the X-ray range to the far infrared but in practice transients are mostly observed in the optical.

Utilizing the models shown in Fig.~\ref{fig_comp_mass_sil_sniin}, and extending the sample to include more dust masses, we show in Fig.~\ref{fig_sniin_extinction} the results of the extinction in the ultraviolet and in the optical (filters $UVW1$ and $V$) for the SN\,IIn model following the presence of dust in the dense shell. As expected, the extinction is greater with increasing dust mass. This extinction is, however, greater in the optical because optical photons arise primarily from within the dense shell whereas ultraviolet photons escape primarily from the outer parts of the dense shell (even in the dust-free case). For a moderate dust mass of a few 10$^{-4}$\,\msun, the  extinction is about 1\,mag in the ultraviolet and optical, corresponding to a flux reduction by a factor 2.5. This extinction can thus impact our inference of the strength of the interaction as well as the emission arising from the inner ejecta (from which one estimates the amount of \nifs, or O, and potentially the progenitor mass).

\begin{figure}
\centering
 \includegraphics[width=0.49\textwidth]{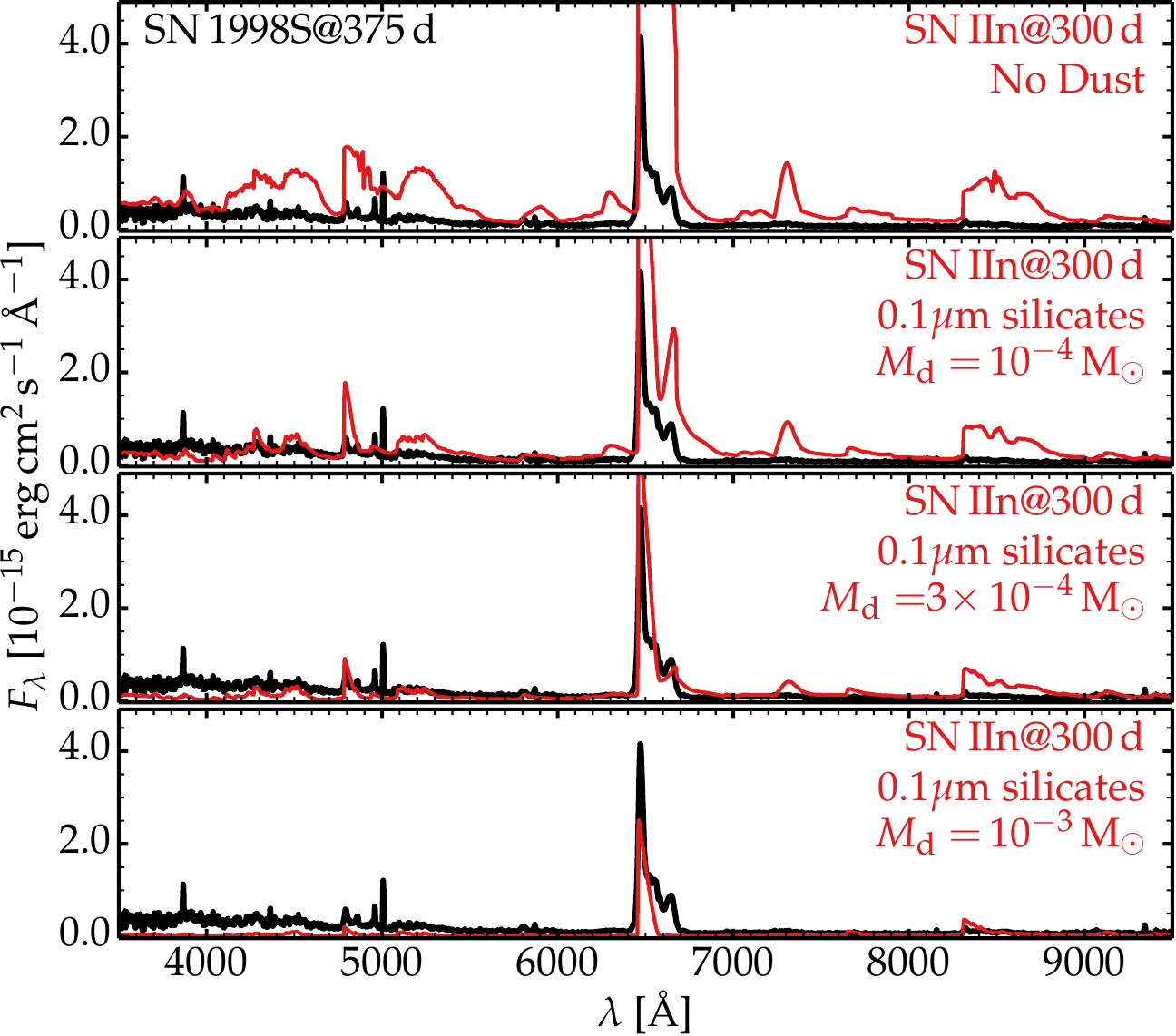}
\caption{Comparison of the optical spectrum of SN\,1998S at 375\,d (black) with synthetic spectra of the SN\,IIn model at 300\,d endowed with different amounts of dust from zero to 0.001\,\msun\ (red). The observed spectrum has been corrected for redshift and reddening. The model spectra have been scaled to the adopted distance of 17\,Mpc to SN\,1998S. When included, the dust is chosen to be 0.1\mic\ silicate grains and located in the outer-ejecta dense shell.
\label{fig_98S_sil}
}
\end{figure}

Finally we show, in Fig.~\ref{fig_98S_sil}, a comparison of a selection of the above models with the observations of SN\,1998S at 375\,d after explosion (there is a slight mismatch in post-explosion epoch but we are mostly interested here in the qualitative aspects). The data are from \citet{leonard_98S_00}, from where we also adopt the reddening $E(B-V)$ of 0.23\,mag, the distance of 17\,Mpc, and a recession velocity of 840\,\kms. With the adopted absorbed shock power of $2\times10^{41}$\,\ergs, the predicted model spectrum exhibits a global flux offset throughout the optical with an additional discrepancy in the strength and morphology of the H$\alpha$ profile (i.e., strong and broad in the dust-free model but strongly asymmetric in the observations). With the introduction of 10$^{-4}$\,\msun\ of dust in the dense shell, these offsets are reduced. Here, a good match is obtained for a dust mass of $3\times10^{-4}$\,\msun, whereas the extinction and blue-red asymmetry is too great for a dust mass of 10$^{-3}$\,\msun. These results imply the presence of dust in SN\,1998S but also indicate that dust should be accounted for when estimating the magnitude of the power source and the associated CSM density (or mass loss rate).

A further implication is that dust in the dense shell contributes to attenuating the emission from the metal-rich inner ejecta regions (e.g., \oidoub\ or \caiidoub, which are strong coolants of the O-rich and Fe-rich gas), making the SN\,IIn appear as if there was less, and potentially no O or \nifs\ ejected in the explosion. This aspect is particularly important for understanding the progenitors and the explosion characteristics of interacting SNe. A dust-attenuated metal-rich inner ejecta could thus be incorrectly interpreted as arising from a low-mass massive-star progenitor (i.e., one with a low-mass He core), perhaps combined with negligible explosive nuclesynthesis. Such properties might then open the possibility that the transient was not a terminal explosion, something that in fact plagues numerous transients with strong signatures of interaction (see, e.g., SN\,2009ip and \citealt{pastorello_09ip_13}, \citealt{margutti_09ip_14}).

\section{Dust in the SN\,IIP/CSM model and comparison to SN\,2017eaw at 900\,d}
\label{sect_sniipcsm}

\begin{figure}
   \centering
    \begin{subfigure}[b]{0.45\textwidth}
       \centering
       \includegraphics[width=\textwidth]{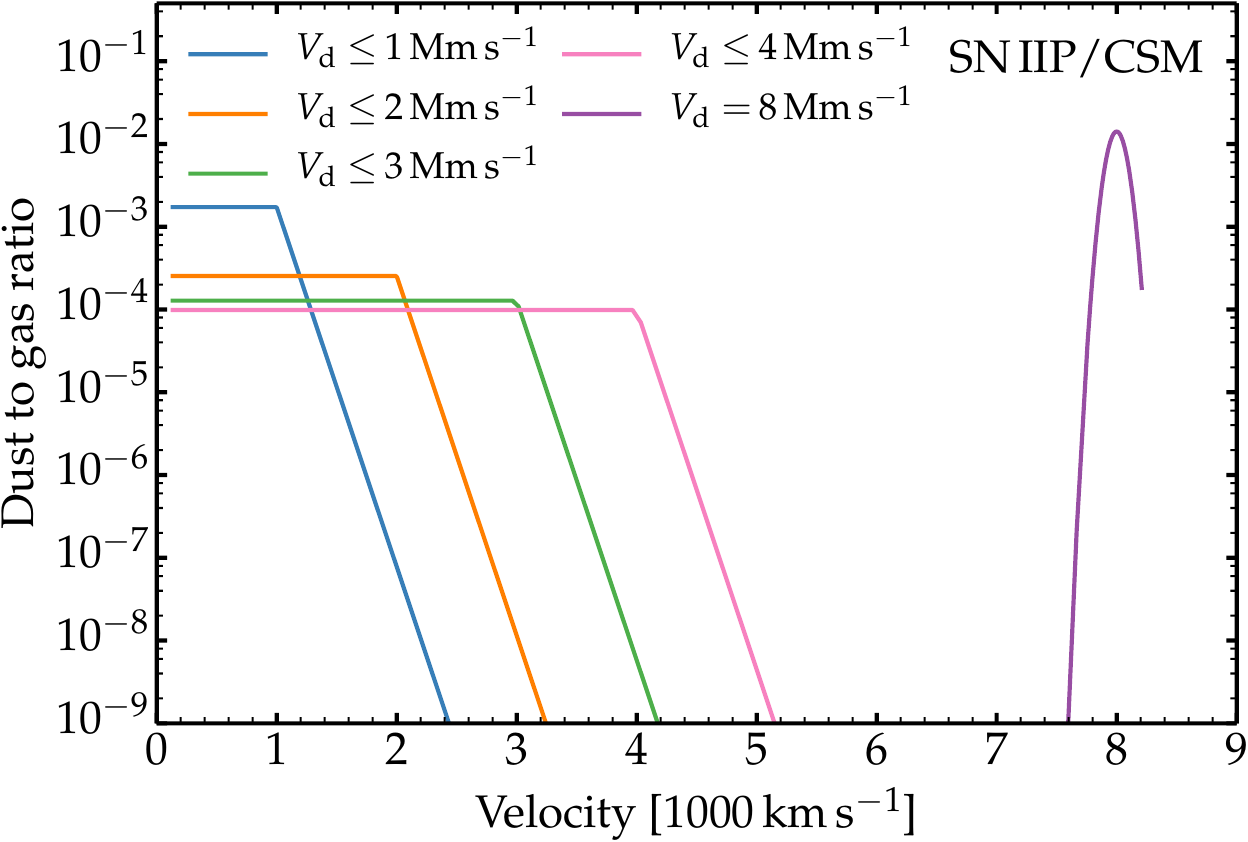}
    \end{subfigure}
    \hfill
    \centering
    \begin{subfigure}[b]{0.45\textwidth}
       \centering
      \includegraphics[width=\textwidth]{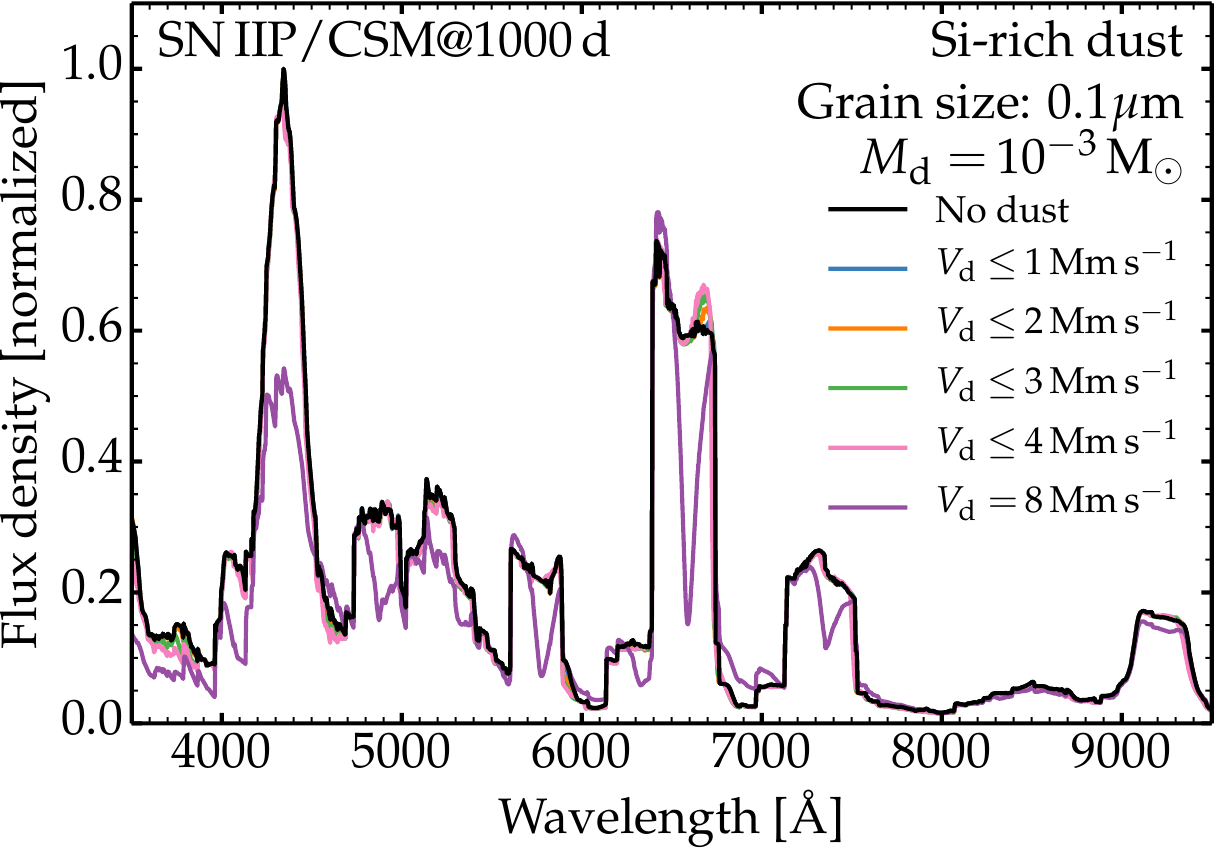}
    \end{subfigure}
    \hfill
    \centering
    \begin{subfigure}[b]{0.45\textwidth}
       \centering
      \includegraphics[width=\textwidth]{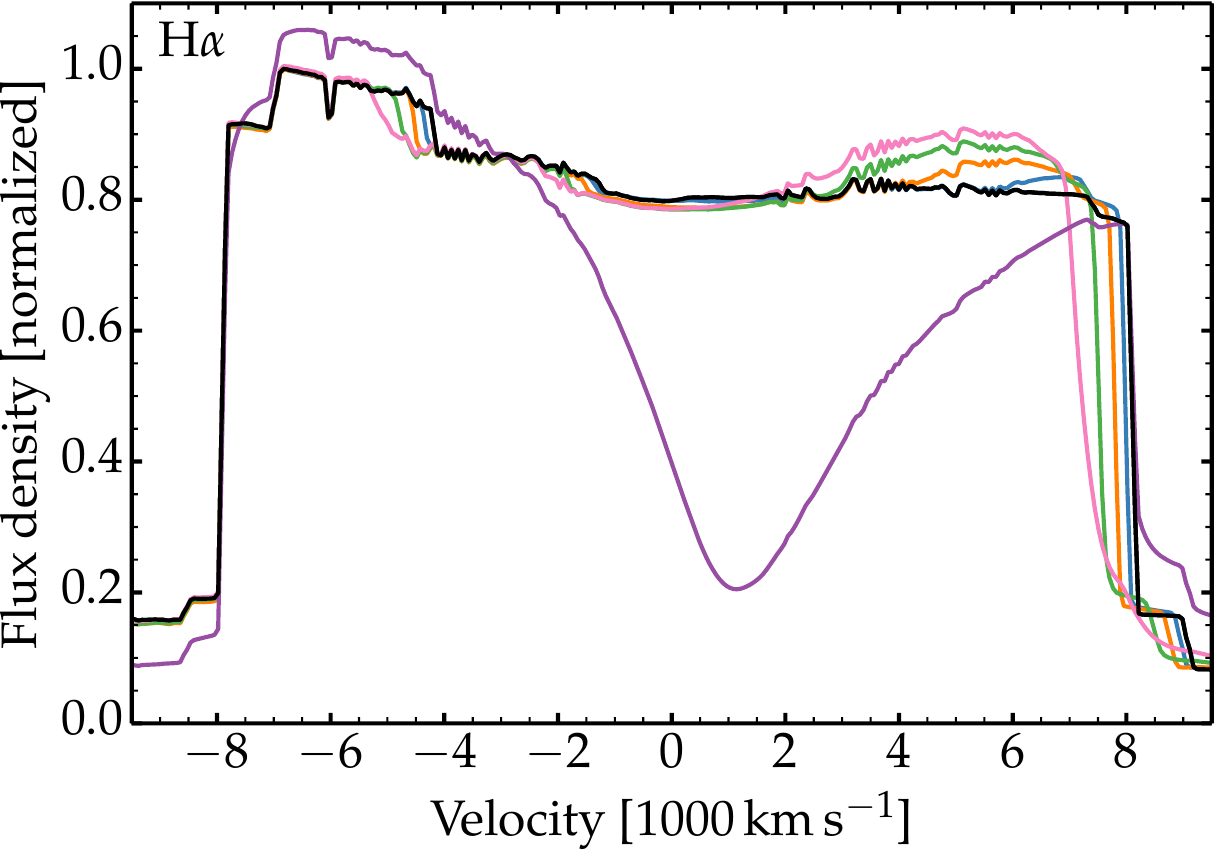}
    \end{subfigure}
\caption{Same as Fig.~\ref{fig_comp_loc_sil_sniin} but now for the SN\,IIP/CSM model at 1000\,d. The outermost location chosen for the dust corresponds to the outer-ejecta dense shell at 8000\,\kms. Some numerical noise affects this simulation because of the rapid variation in properties in the fast, narrow dense shell. [See Section~\ref{sect_sniipcsm} for discussion.]
\label{fig_s15p2_comp_m1em3_vupdust}
}
\end{figure}

In this section, we consider the impact of dust in a SN\,II at 1000\,d that is powered by both radioactive decay in the inner ejecta ($L_{\rm decay,abs}=3.2\times10^{37}$\,\ergs) and by interaction with the progenitor wind in the outer ejecta ($L_{\rm sh,abs}=10^{40}$\,\ergs) -- the SN\,IIP/CSM model is taken from \citet{dessart_late_23}. Figure~\ref{fig_s15p2_comp_m1em3_vupdust} is an analog of Fig.~\ref{fig_comp_loc_sil_sniin} and shows the impact on the optical spectrum of the SN\,IIP/CSM model at 1000\,d for dust present in various locations within the inner ejecta or exclusively in the dense shell at 8000\,\kms. For this illustration, we use 0.1\mic\ silicates and a fixed total dust mass of 0.001\,\msun. The dust has essentially no impact on the emergent spectrum unless it is located in the dense shell. This arises because 99.7\,\% of the emission in this model comes from the dense shell (this contrast is smaller though still large when considering only the optical) and because the inner-ejecta regions subtend a small angle as viewed from the outer dense shell (their ability to occult the backside of the emitting dense shell is small). With the adopted dust mass of 0.001\,\msun, the dust-to-gas ratio in the dense shell peaks at values greater than 0.01, which is uncomfortably high given the solar composition in those H-rich layers. But only for such conditions is the dust optical depth approaching unity in the dense shell -- for a lower dust mass of 10$^{-4}$\,\msun\ the impact on the spectrum is negligible. In SNe that have expanded for several years, any dust in the fast, outer ejecta is most likely optically thin and thus has little impact on the spectrum (unless some interaction has occurred and led to the formation of a massive, dense shell in the outer ejecta -- see Section~\ref{sect_sniin}).

When the dust is located in the dense shell and is optically thick (i.e., here if the dust mass is about 0.001\,\msun), all emission lines suffer some extinction, and that extinction is maximum for the region in the limbs of the shell (the dust optical depth is greater for those impact parameters) and causes a pronounced dip near the rest wavelength. This is in part an artifact of our strict assumption of spherical symmetry. If instead the dense shell was allowed to break up laterally, the magnitude of that central dip would be reduced \citep{flores_shell_22}, although that would also tend to diminish optical-depth effects associated with the dust.

\begin{figure}
\centering
\includegraphics[width=0.49\textwidth]{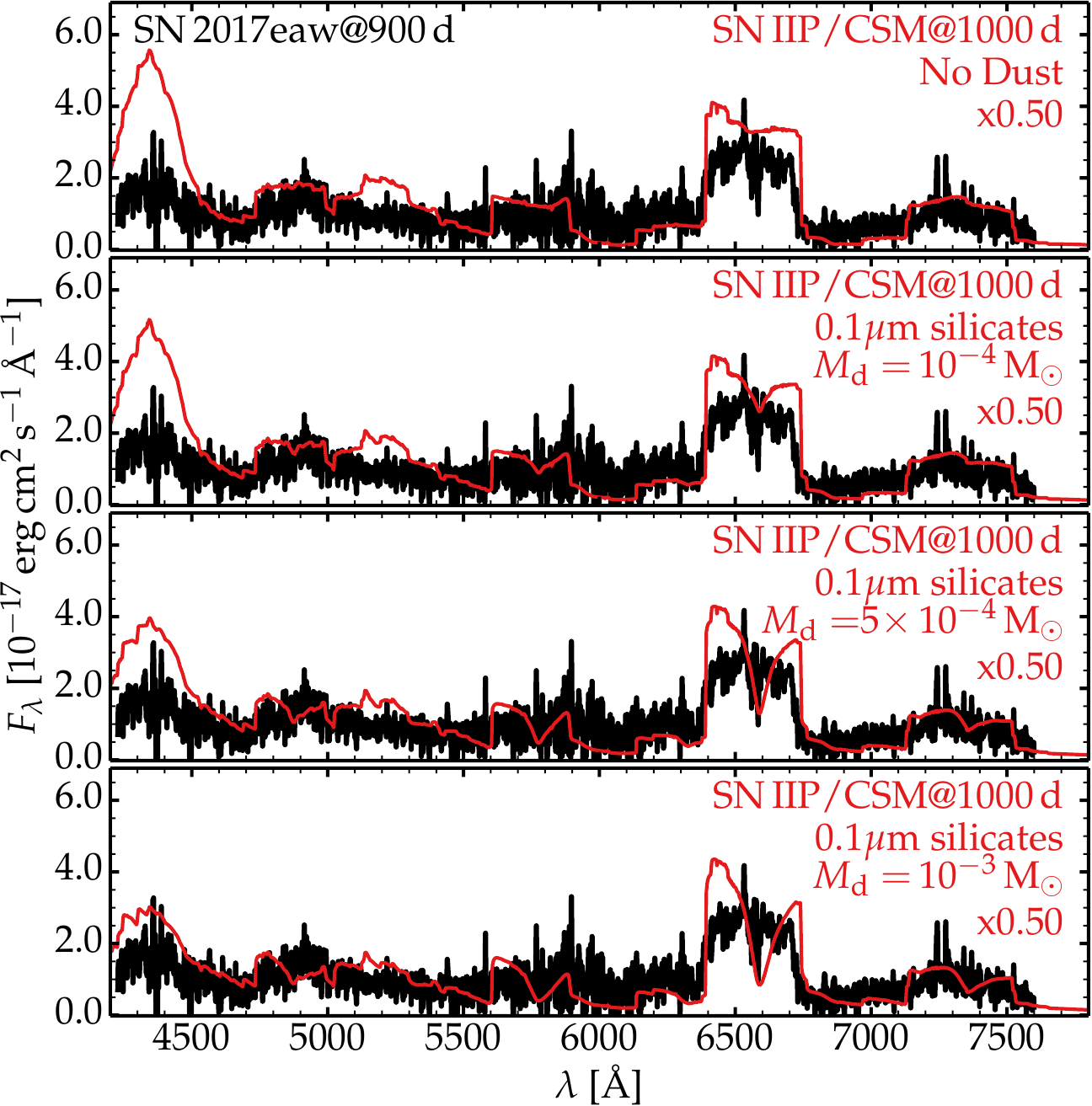}
\caption{Comparison of the optical spectrum of SN\,2017eaw at 900\,d (black) with synthetic spectra of the SN\,IIP/CSM model at 1000\,d with different assumptions on the dust properties (red). The observed spectrum has been corrected to match the $R$-band photometry of \citet{weil_17eaw_20}, and then corrected for redshift and reddening. The model spectra have been scaled to the SN\,2017eaw distance, with an additional scaling of 0.5. When included, the dust is chosen to be 0.1\mic\ silicate grains and located in the outer dense shell at 8000\,\kms.
 [see Section~\ref{sect_sniipcsm} for discussion.]
\label{fig_sniipcsm_17eaw}
}
\end{figure}

Figure~\ref{fig_sniipcsm_17eaw} compares a sample of our dusty SN\,IIP/CSM models with the observations of SN\,2017eaw at 900\,d. The data are from \citet{weil_17eaw_20} and we adopt the distance of 7.73\,Mpc, the reddening $E(B-V)=0.3$\,mag and the redshift of 0.00013 from \citet{vandyk_17eaw_19}. The observations show a broad, boxy, and roughly symmetric H$\alpha$ profile with little emission in other lines or in the continuum. It is not clear that any dust is needed to reproduce this optical spectrum (unlike for SN\,1998S at 375\,d) and indeed our models with various amounts of dust (in the dense shell) yield a rough match to the observations. Using optical and infrared observations of SN2017eaw at $\sim$\,5 years (thus $\sim$\,2.5\,yr later than the present comparison), \citet{shahbandeh_jwst_23} infer the presence of several 10$^{-4}$\,\msun\ of dust in the ejecta regions below 3000\,\kms.

\section{Dust in the SN\,II model and comparison to SN\,1987A at 700d}
\label{sect_snii}

We now turn to the more universal configuration of a standard, noninteracting SN\,II model powered exclusively through radioactive decay. The model age is 700\,d and is powered with $L_{\rm decay,abs}=8.3\times10^{38}$\,\ergs. We do not repeat the explorations of the previous sections and consider only the possibility of the presence of dust in the inner ejecta, as expected theoretically (see, e.g., \citealt{sarangi_dust_22}) and inferred observationally (see, e.g., \citealt{lucy_dust_89,bevan_barlow_16}).

\begin{figure}
\includegraphics[width=0.49\textwidth]{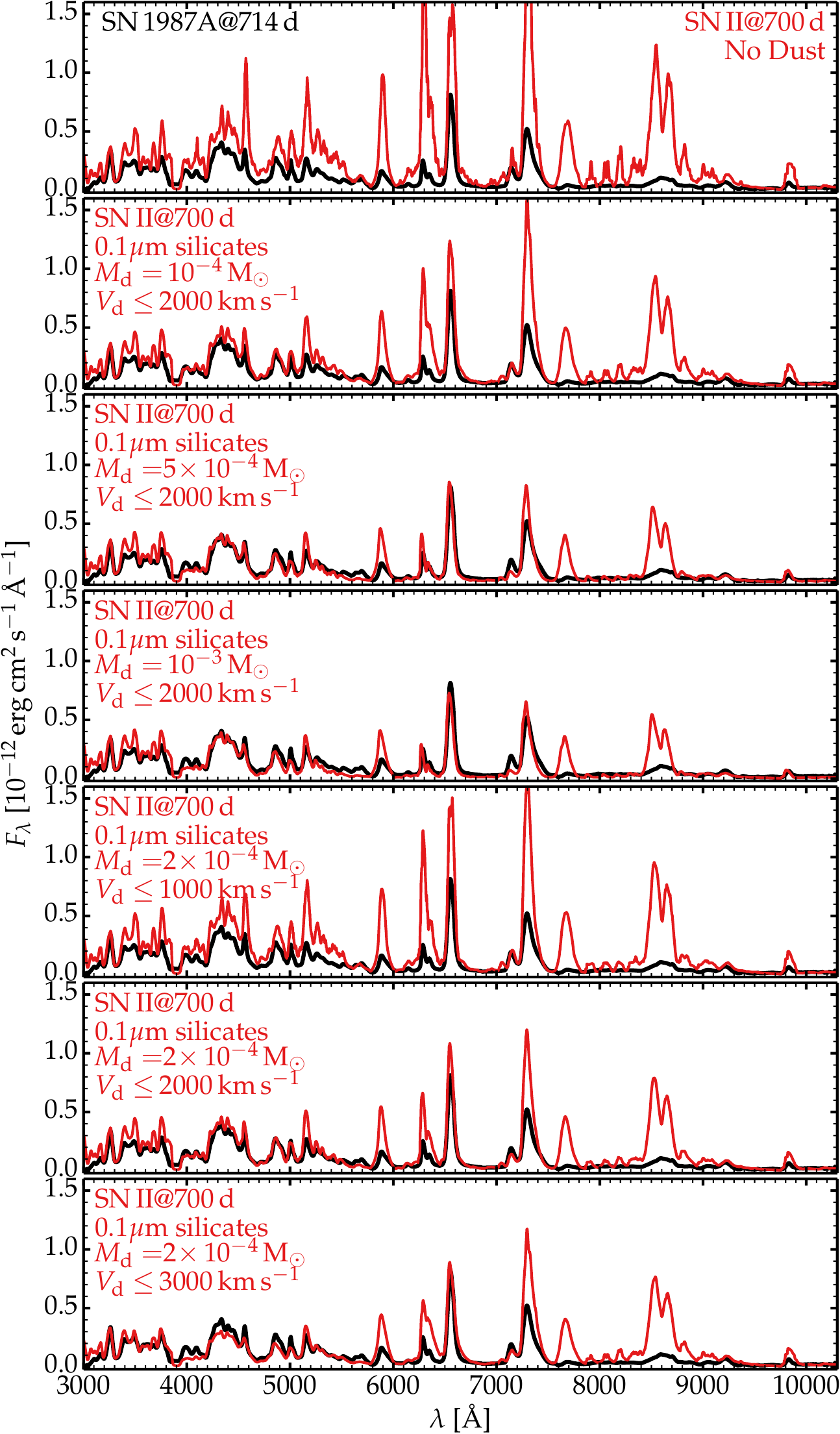}
\caption{Comparison of the optical spectrum of SN\,1987A at 714\,d (black) with synthetic spectra of the SN\,II model at 700\,d with different assumptions on the dust properties (red). The observed spectrum has been corrected for redshift and reddening. The model spectra have been scaled to the SN\,1987A distance. When included, the dust is chosen to be 0.1\mic\ silicate grains and located in the inner metal-rich ejecta. [See Section~\ref{sect_snii} for discussion.]
\label{fig_snii_87A}
}
\end{figure}

Figure~\ref{fig_snii_87A} shows the results of a comparison between the observations of SN\,1987A at 714\,d and the SN\,II model with various choices of dust. The data are from \citet{phillips_87A_90} and we have adopted a distance of 49.59\,kpc \citep{pietrzynski_lmc_dist_19}, a reddening $E(B-V)=0.15$\,mag, and a redshift of 0.00096. The models consider the possible presence of dust within 1000, 2000, and 3000\,\kms, as well as dust masses between 10$^{-4}$ and 10$^{-3}$\,\msun. The dust-free case is shown at top for comparison.

Our model has a \nifs\ mass of 0.063\,\msun\ and thus close to the inferred value of 0.07\,\msun\ for SN\,1987A \citep{sn1987A_rev_90}. However, in the absence of dust, our model, reddened and scaled to the distance of SN\,1987A is too bright (top panel of Fig.~\ref{fig_snii_87A}), although it matched well the observations of SN\,1987A at 350\,d \citep{dessart_sn2p_21}. As discussed in \citet{lucy_dust_89}, dust extinction is likely the cause of the offset in brightness. Introducing dust with a mass of $\sim$\,10$^{-4}$\,\msun\ reduces this photometric offset. A dust mass of 10$^{-4}$\,\msun\ yields a good match in the blue part of the optical but has an insufficient effect on most emission lines -- the same is true if the dust is placed too deep in the ejecta (e.g., below 1000\,\kms). Raising the dust mass to $5-10\times10^{-4}$\,\msun\ yields a satisfactory match throughout the optical if the dust encompasses the entire metal-rich inner ejecta, which extend out to about 2000\,\kms. In that case, most emission lines exhibit the same strength as observed, with the notable exceptions of \nad, \kidoub, and the Ca\two\ near-IR triplet. This offset may be related to the adopted solar-, rather than LMC-metallicity of the progenitor model, or it may be indicative that a small amount of dust is also present further out in the ejecta. Interestingly, we see that in the best fitting models ($M_{\rm d}=5-10\times10^{-4}$\,\msun\ and $V_{\rm d}\le2000$\,\kms), the flux ratio between \oidoub\ and H$\alpha$ is strongly altered with the introduction of dust. This arises because \oidoub\ forms exclusively within the metal-rich inner ejecta where we have introduced the dust, whereas H$\alpha$ forms both in the inner ejecta (because of previous inward mixing of H to low velocities) as well as above those inner metal-rich regions, which we adopted to be free of dust. There is thus a differential effect of dust attenuation, biased against the metal rich inner regions where lines like \oidoub\ and \caiidoub\ preferentially form.

Overall, our findings are in agreement with the model results for dust formation in Type II SN ejecta by \citet{sarangi_dust_22} who infers a dust mass of $7\times10^{-4}$\,\msun\ below 1500\,\kms\ at 700\,d after explosion. They also agree with the silicate dust mass of $3.2 \times 10^{-4}$\,\msun\ inferred by \citet{lucy_dust_89} from the modeling of a selection of optical line profiles in SN\,1987A at 775\,d. Similarly, \citet{bevan_barlow_16} estimated the dust mass in SN\,1987A at 714\,d (and out to about 10\,yr) by modeling the asymmetry of the H$\alpha$ and \oidoub\ emission profiles. Adopting carbonaceous dust grains with a size between 0.35 and 3.5\mic, they infer a dust mass roughly in the range between 10$^{-5}$ and 10$^{-3}$\,\msun. This is rather a large range and suggests a sizable uncertainty. Differences in results obtained for H$\alpha$ and \oidoub\ can in part arise from the fact that \oidoub\ forms systematically deeper and in more dust-rich regions (and thus more sensitive to the presence of dust) than H$\alpha$. Our value of $5-10\times10^{-4}$\,\msun\ of dust (although for 0.1\mic\ silicate grains) falls within their proposed range at that time (\citealt{bevan_barlow_16} inferred an increase in dust-mass by nearly a factor of a hundred between two and ten years after explosion, which is comparable to the model predictions of \citealt{sarangi_dust_22}).

\section{Conclusions}
\label{sect_conc}

A module for the treatment of dust has been added to the non-LTE time-dependent radiative transfer code \cmfgen\ \citep{HD12}. At present, dust is introduced in the final calculation of the model spectrum with \cmfflux\ and is thus not fully coupled to the gas. One thus specifies the type, amount, spatial distribution, and (single) temperature of the dust. For this work, we have considered 0.1\mic\ silicate grains and a dust temperature of several 100\,K but we varied the mass and distribution of the dust within the ejecta. Where the dust is present, its radial variation follows the mass density profile. Although some of our models employ clumping in the form of a radial compression, we ignore any additional compression in the lateral direction, which would tend to reduce the effective dust opacity. This inherent complication of SN ejecta structure leads to an underestimate of inferred dust masses. We also considered a representative sample of ejecta in order to cover interacting (such as SN\,1998S at $\sim$\,1\,yr and SN\,2017eaw at $\sim$\,3\,yr) as well as noninteracting SNe II (i.e., SN\,1987A at $\sim$\,2\,yr). The influence of dust on the  spectrum depends strongly on the location of the dust relative to the regions that give rise to the ``dust-free'' SN spectrum.

To study the influence of dust on the observed spectrum we utilized two approaches -- a classic solution of the radiative-transfer equation and a Monte-Carlo approach. To test the influence of the phase function we considered both isotropic scattering and highly-anisotropic scattering (specified using the Henyey-Greenstein phase function with $g=0.8$).  Comparisons between the two codes showed good agreement when we assumed either isotropic of anisotropic scattering. This a rigorous accuracy test since the two codes use very different solution techniques. For the present case either solution approach works but for arbitrary phase functions and polarization calculations the Monte-Carlo code has distinct advantages. Given the uncertainties/freedoms in the grain size distributions, dust location and dust mass utilizing isotropic scattering generally provides an adequate approach for modeling integrated spectra of SNe that are affected by dust.

In the SN\,IIn model tailored for SN\,1998S at about one year post explosion, we find that only the dust from within the dense shell at $\sim$\,5000\,\kms\ can impact the model spectrum in the ultraviolet and in the optical. This arises because essentially all the radiation originates from within the dense shell and because any dust present in the inner ejecta holds too small a subtended angle to affect the emission from the dense shell. Dust affects all emission line profiles by causing a flux deficit redward from line center (i.e., emission from the receding part of the dense shell), but also a global attenuation of the flux at all wavelengths. Paradoxically, this attenuation by dust is greater in the optical, which forms from within the dense shell where the dust is located, and lower in the ultraviolet, which forms in the external part of the dense shell (there is already strong ultraviolet attenuation in the dense shell without dust).  We find that 0.0003\,\msun\ of dust can explain the blue-red asymmetry of H$\alpha$ and the overall optical spectrum of SN\,1998S at one year. This level of dust causes an extinction of $\sim$\,1.8\,mag in the $V$ band and of $\sim$\,1.0\,mag in the $UVW1$ band.

In the Type IIP SN model with weak, late-time interaction, the outer dense shell is less massive and is located at much larger velocity. Only for very large and probably unrealistic dust-to-gas ratios (i.e., implying a near complete use of the primordial metals from this H-rich gas to form the dust) can dust impact line profiles. This comes mostly as a central flux deficit and blue-red asymmetry on H$\alpha$, H$\beta$ etc. The quasi flat-topped H$\alpha$ profile in SN\,2017eaw at 900\,d is compatible with a dust mass of several 0.0001\,\msun. Dust formed from within the inner, metal-rich ejecta has no impact on that broad H$\alpha$ line profile. 

In nature, the power absorbed in the dense shell and reprocessed as optical photons may vary in magnitude -- the value of 10$^{40}$\,\ergs\ used here and in \citet{dessart_late_23} may be on the high side. That shock power could be smaller and closer to the decay power absorbed in the inner ejecta regions at those late times. In that case, the observation of broad lines arising from the dense shell and the absence of narrower lines arising from the inner ejecta would require the presence of dust in the inner ejecta in order to obscure those regions to an external observer -- even though no blue-red asymmetry would be visible from the observed, broad and boxy emission line profiles arising from the dense shell.  

In the noninteracting Type II SN model, the introduction of dust in the inner, metal-rich ejecta causes a differential attenuation of the ejecta emission. Strong forbidden lines such as \oidoub\ and \caiidoub, which form nearly exclusively in the O-rich and Fe/Si-rich regions, respectively, are strongly affected by dust located within 2000\,\kms. In contrast, the emission from H-rich material which causes the Fe\two\ emission below 5500\,\AA\ or H$\alpha$ is only partially affected because a large fraction of that power arises from exterior regions,  outside 2000\,\kms: This emission appears stronger relatively to \oidoub\ and \caiidoub\ which are quenched. This effect counteracts the strengthening of these lines that occurs in the absence of dust, inhibiting the evolution of the spectrum past 500\,d (i.e., the spectrum of SN\,1987A at 700\,d is analogous to that at 350\,d, with the exception of the stronger Fe\two\ emission below 5500\,\AA).

By coupling dust with the full mixture of atoms and ions within the ejecta, we could solve for the dust temperature at all depths in \cmfgen, rather than prescribe a dust temperature (and the same one at all depths) as currently done. The greater physical consistency would allow for a proper assessment of infrared emission from dusty SNe, although it requires the complete and challenging modeling of the ejecta and the various power sources involved. A more sophisticated treatment of the dust-grain size distribution (e.g., with a power law), the dust chemistry (e.g., treating multiple types of dust), or the dust spatial distribution (e.g., allowing its presence in multiple regions) would also improve the realism of our simulations. This is left to future work.

\begin{acknowledgements}
LD thanks the Pittsburgh particle-physics, astrophysics, and cosmology center for financial support during a 2024 winter visit to the University of Pittsburgh. D.J.H. gratefully acknowledges support through NASA astro-physical theory grant 80NSSC20K0524. This work was granted access to the HPC resources of TGCC under the allocation 2024 -- A0170410554 made by GENCI, France. This research was supported in part by the University of Pittsburgh Center for Research Computing and Data, RRID:\,SCR\_022735, through the resources provided. Specifically, this work used the H2P cluster, which is supported by NSF award number OAC--2117681. This research has made use of NASA's Astrophysics Data System Bibliographic Services.
\end{acknowledgements}

\appendix

\section{Dust Scattering and the Henyey-Greenstein Phase Function}
 \label{append_HG}

The Henyey-Greenstein (HG) phase function is
\begin{equation}
P(\Theta)= {1 \over 4 \pi}  {1-g^2 \over (1+ g^2 -2g \cos\Theta)^{3/2}}
\label{eq_hg}
\end{equation}

\noindent
where $\Theta$ is the scattering angle and is defined by $\cos \Theta =\vec{n_i} \cdot \vec{n_s}$ where $\vec{n_i}$ and $\vec{n_s}$ refer to the unit vectors
describing the direction of the incident and scattered rays and $ -1 < g < 1$. The  HG phase function
was introduced by \citet{henyey_greenstein_41} to provide a simple, one parameter, formula that could be used to describe different types of anisotropic scattering. In many cases it provides an excellent first-order approximation to describe anisotropic scattering by dust, but for more accurate descriptions other approximate formulae have been proposed (e.g., \citealt{baes_dust_22}).

To use the HG functions in a radiative transfer code we need to rewrite it in terms of $(\theta_i, \phi_i)$ and $(\theta_s,\phi_s)$ which
describe the orientation of the incident and scattered rays relative to the local radius vector.
Using
\begin{equation}
\vec{n_i}=\cos \phi_i \sin \theta_i \,\vec i+  \sin\phi_i \sin \theta_i \, \vec j + \cos \theta_i \,\vec k
\end{equation}
and a similar expression for $\vec{n_s}$
\noindent
we have (after some simple manipulations)
\begin{eqnarray}
\cos \Theta &=&\sin \theta_i  \sin \theta_s \cos(\phi_s-\phi_i) + \cos \theta_i \cos \theta_s \\
&=&\mu_i \mu_s +  \sqrt{1-\mu_i^2} \, \sqrt{1-\mu_s^2}  \cos(\phi_s-\phi_i)
\end{eqnarray}
and
\small{
\begin{eqnarray}
P(\Theta)&=&  {1 \over 4 \pi} {1-g^2 \over [1+ g^2 -2g \mu_i \mu_s-  w(g,\mu_i,\mu_s) \cos(\phi_s-\phi_i)]^{3/2}} \\
&=&  { c(g,\mu_i, \mu_s)  \over [1 -a(g,\mu_i, \mu_s) \cos(\phi_s-\phi_i)]^{3/2} }
\label{eq_theta}
\end{eqnarray}   }
where
\begin{equation}
w(g,\mu_i,\mu_s)= 2g \sqrt{1-\mu_i^2}\sqrt{1-\mu_s^2} \,\,,
\end{equation}
\begin{equation}
c(g,\mu_i, \mu_s) = {1 \over 4\pi} {1-g^2 \over (1+ g^2 -2g \mu_i \mu_s)^{3/2}}
\end{equation}
and
\begin{equation}
a(g,\mu_i ,\mu_s)= { 2g \sqrt{1-\mu_i^2}\sqrt{1-\mu_s^2}  \over 1+ g^2 -2g \mu_i \mu_s }
\end{equation}
with  $ -1 \le a \le 1$.

As we are considering dust scattering in spherical geometry the radiation field is independent of azimuth. Thus we only need to
consider the azimuthally averaged HG phase function. To obtain the azimuthal average we need to average the term
\begin{equation}
{1 \over [1 -a \cos(\phi_s-\phi_i)]^{3/2} } \,\,
\end{equation}
over $\phi$. Further, we can treat $a$ as a constant since it does not depend on $\phi$. We define $A$ by
\begin{equation}
 A(a) = \int_0^{2\pi} {1 \over (1- a \cos \phi)^{3/2}} \, d\phi
\end{equation}
so that
\begin{equation}
\tilde  P(g,\mu_i,\mu_s)=c(g,\mu_i, \mu_s) A(a)\,\,.
\end{equation}

\noindent
To perform the integration over $\phi$ we use Gauss-Legendre quadrature.

At each grid point, the scattered emissivity, $\eta(\mu,g)$,  is found by numerical quadrature.
\begin{eqnarray}
\eta(\mu,g)_s&=&   \sigma_s \int_{-1}^1 \int_{0}^{2\pi} P(g,\mu_s,\mu_i) I(\mu_i) \,d\phi\, d\mu_i \\
&=&\sigma_s \int_{-1}^1 \tilde  P({g,\mu_s, \mu_i}) \,  I(\mu_i) \,d\mu_i  \\
&=& \sigma_s \Sigma_{\mu_i} w_{s}(\mu_i) .  I(\mu_i)
\end{eqnarray}
As $\tilde P(g,\mu_i,\mu_s)$ can vary rapidly with $\mu$ we fit it using a piecewise monotonic cubic. We then integrate the monotonic cubic assuming that $I(\mu)$ is a piecewise linear function of $\mu_i$. More sophisticated integration formulae cannot be used since the $\mu$ are set by our ($p,z$) grid. Assuming
\begin{equation}
	\tilde P(\mu)= c_4 +c_3(\mu-\mu_0)+c_2(\mu-\mu_0)^2+ c_1(\mu-\mu_0)^3
\end{equation}
and
\begin{equation}
	I(\mu)=a+b(\mu-\mu_0)
\end{equation}
we have (writing $\delta_0 = (\mu-\mu_0)$, $\delta_{10} = (\mu_1-\mu_0)$, $I_0\equiv I(\mu_0)=a$, and $I_1\equiv I(\mu_1)=a+b\delta_{10}$)
\begin{eqnarray}
	\int_{\mu_0}^{\mu_1} \bar P(\mu)I(\mu)\,d\mu&=& a\left[ ( c_4 \mu +\frac{1}{2}c_3\delta_0^2 +\frac{1}{3}c_2 \delta_0^3+ \frac{1}{4}c_1\delta_0^4)\right]_{\mu_0}^{\mu_1} \nonumber\\
	&& +  b \left[  \frac{1}{2}c_4 \delta_0^2 +\frac{1}{3}c_3\delta_0^3 +\frac{1}{4}c_2\delta_0^4+ \frac{1}{5}c_1\delta_0^5
	          \right]_{\mu_0}^{\mu_1} \\	         
	&=& a\left[ \delta_0\left(c_4+\frac{1}{2}c_3\delta_0+\frac{1}{3}c_2\delta_0^2+ \frac{1}{4}c_1\delta_0^3 \right) \right]_{\mu_0}^{\mu_1} \nonumber\\
	&& + b\left[ \delta_0^2 \left( \frac{1}{2}c_4  +\frac{1}{3}c_3\delta_0 +\frac{1}{4}c_2\delta_0^2+ \frac{1}{5}c_1\delta_0^3 \right)\right]_{\mu_0}^{\mu_1} \\
&=&af(\mu_1)+bh(\mu_1)\delta_{10} \\
&=& I_0f(\mu_1) +(I_1-I_0)/\delta_{10} h(\mu_1)( \mu_1-\mu_0) \\
&=&I_0[f(\mu_1)-h(\mu_1)] + I_1 h(\mu_1)  \\
&=&w_0 I_0 + w_1 I_1
\end{eqnarray}
where
\begin{equation}
f(\mu_1)=\delta_{10}\left(c_4 + \frac{1}{2}c_3\delta_{10}+\frac{1}{3}c_2\delta_{10}^2+ \frac{1}{4}c_1\delta_{10}^3 \right)\,\,\,
\end{equation}
\begin{eqnarray}
w_1&=&h(\mu_1)  \\
            &=&  \delta_{10}\left(\frac{1}{2}c_4  +\frac{1}{3}c_3\delta_{10} +\frac{1}{4}c_2\delta_{10}^2+ \frac{1}{5}c_1\delta_{10}^3 \right)
\end{eqnarray}
and
\begin{eqnarray}
w_0&=& f(\mu_1)-h(\mu_1) \\
&=&\delta_{10}\left( \frac{1}{2} c_4 + \frac{1}{6}c_3\delta_{10}+\frac{1}{12}c_2\delta_{10}^2+ \frac{1}{20}c_1\delta_{10}^3 \right)
\end{eqnarray}

\noindent
In the radiative transfer code we can precompute the quadrature weights since they only depend on the adopted grid, and are independent of frequency. In the Monte-Carlo code the treatment of dust scattering is much more simple. We simply tabulate the HG phase function as a function of scattering angle. Using this table we can easily select a new direction for the scattered beam of photons.

\section{More on dust calculations}
 \label{append_dc}

In this appendix we provide some simple test calculations that illustrate both the accuracy of the profile calculations and the influence of anisotropic scattering. The dust distribution is Gaussian with a FWHM of 166\,\kms, and is centered on the dense shell at 5000\,\kms. The dust is assumed to be silicate grains with a size of 0.1\mic, and the total dust mass is $10^{-4}\,$\msun.  For all transfer calculation we assumed a fixed Doppler width of 10\,\kms\ -- values of 5\,\kms\ and 20\,\kms\ yield similar profiles. To describe the anisotropic scattering we used  the HG phase function with $g=0.8$. With this value of $g$, the scattering is strongly forward-peaked and thus provides a strong test of the solution techniques. Actual values of $g$ are likely to be lower. For example, in their study of an intermediate-latitude diffuse cloud, \citet{zhang_dust_23} found  that the scattering in the optical red and green passbands was reasonably well described using a HG phase function with $g\approx 0.53$.

\begin{figure}[h]
\centering
\includegraphics[width=0.49\textwidth]{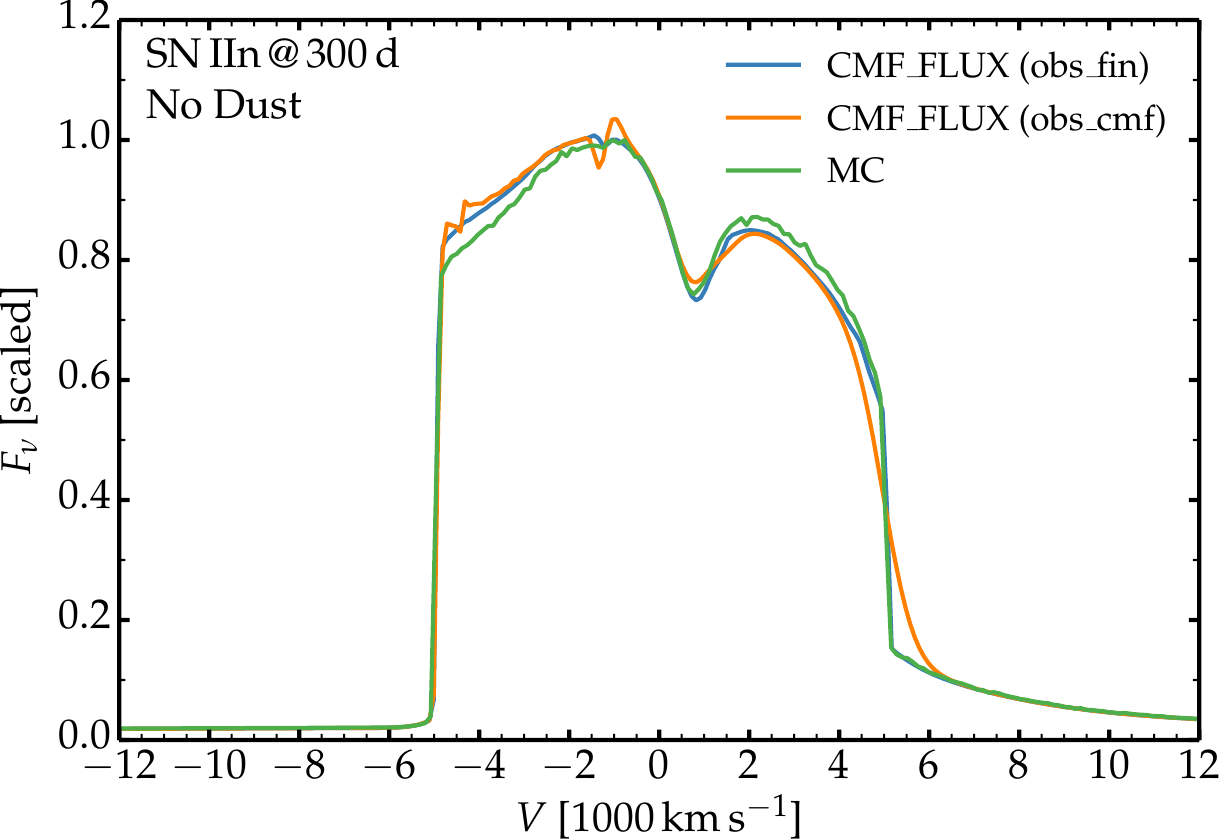}
\caption{Comparison in the SN\,IIn model at 300\,d of the H$\alpha$ profile produced by the MC and \cmfflux\ calculations when we ignore dust. For \cmfflux, we show both the results from the CMF and from the observer's frame calculations.
\label{fig_comp_mc_cmfgen_no_dust}
}
\end{figure}

In Fig.~\ref{fig_comp_mc_cmfgen_no_dust} we show the observed H$\alpha$ line profile computed with \cmfflux\ in the CMF or in the observer's frame (obs\_cmf and obs\_fin), and computed using the Monte-Carlo code -- the dust is ignored in this first test.  Excellent agreement between the three sets of calculations is seen. The CMF line profile shows a slight smoothing on the red-side due to the effect of numerical diffusion discussed earlier in Section~\ref{sect_cmfgen}. The red side is slightly weaker than the blue side due to the influence of electron scattering.

\begin{figure}[h]
\centering
\includegraphics[width=0.49\textwidth]{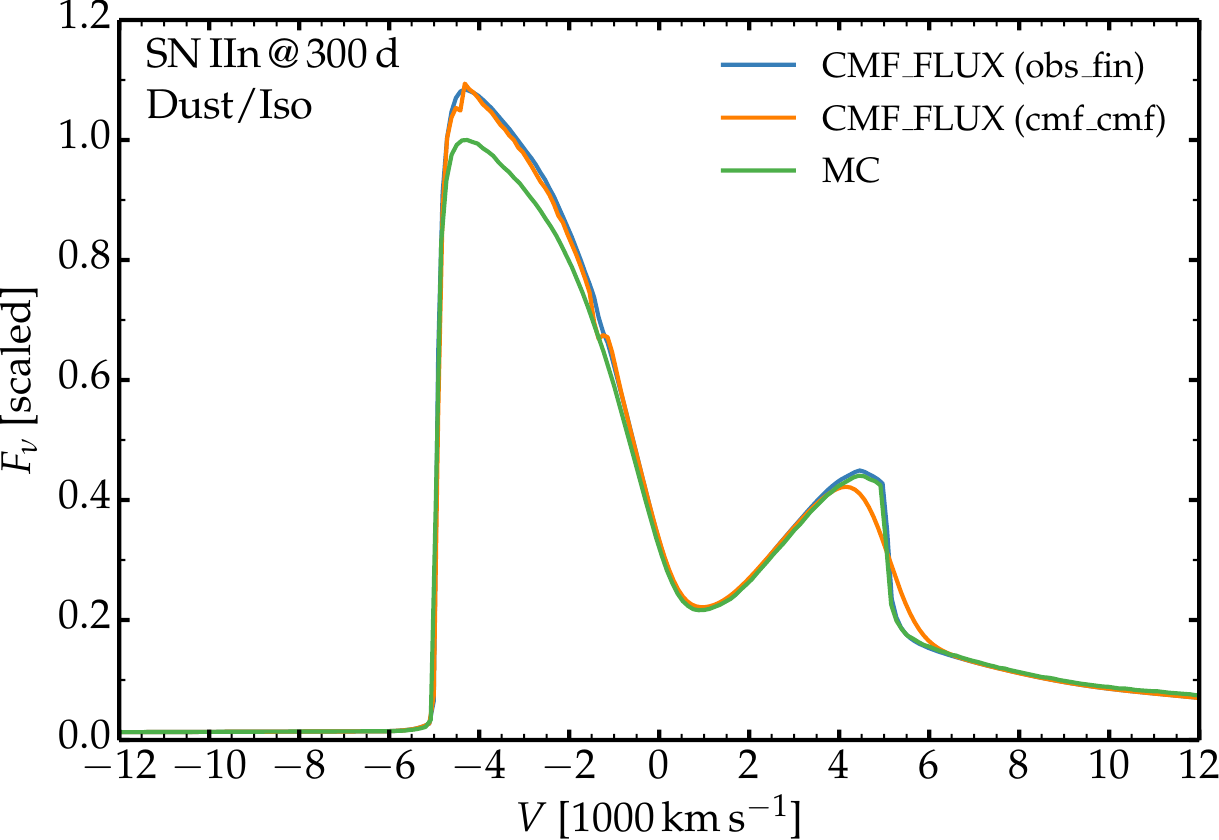}
\caption{Comparison in the SN\,IIn model at 300\,d of the H$\alpha$ profile between the MC code and \cmfflux\ when we assume isotropic scattering. All three profiles show similar agreement to that obtained in the absence of dust (see Fig.~\ref{fig_comp_mc_cmfgen_no_dust}). The red component of the CMF calculation (i.e., obs\_cmf) shows some rounding due to numerical diffusion in frequency space.
\label{fig_comp_mc_cmfgen_gp0}
}
\end{figure}

In Fig.~\ref{fig_comp_mc_cmfgen_gp0} we show the influence of isotropic dust scattering. Due to dust scattering the red peak is strongly suppressed.  Slight differences with the Monte-Carlo approach are seen -- these most likely arise from the very different numerical approaches and the use of the Sobolev approximation in the Monte-Carlo code.

\begin{figure}[h]
\centering
\includegraphics[width=0.49\textwidth]{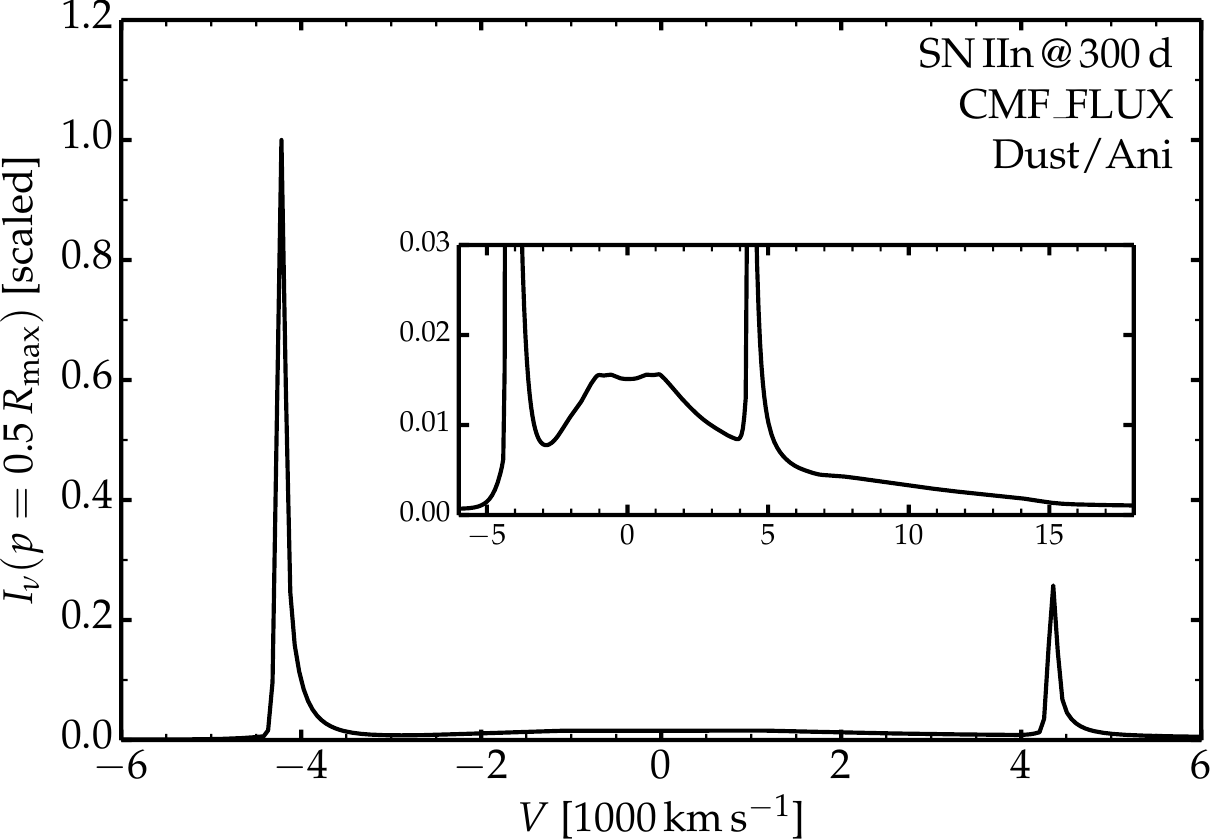}
\caption{Illustration of the emergent mean intensity in the SN\,IIn model at 300d and assuming anisotropic dust scattering. We show the quantity $I(p)$ for the impact parameter $p=$\,0.5\,$R_{\rm max}$. Two narrow H$\alpha$ components, due to the dense shell, are apparent, one redshifted and one blueshifted. The  component at large velocities ($V > 5000$\,\kms) seen in the inset is due to dust scattering while the component centered on 0\,\kms\ is due to ejecta emission.
\label{fig_comp_mc_cmfgen_gp8_ip}
}
\end{figure}

For insight into the broad boxy profiles arising from the dense shell we provide two additional illustrations. Figure~\ref{fig_comp_mc_cmfgen_gp8_ip} shows the observed H$\alpha$ profile for one line of sight. From the figure we see that there are two narrow components that represent the H$\alpha$ emission from the dense shell. Because the shell is narrow we only see emission at two distinct velocities, with their width set primarily by the velocity width of the shell. In the inset, we see the influence of the dust scattered component -- it extends to over 15\,000\,\kms\ on the red side. The broad component centered on 0\,\kms\ is primarily emission from the inner SN ejecta. In the case of a narrow dense shell, the emission (projected on the sky) at a given wavelength arises from a circular ring on the sky. By contrast, the line emission at a given wavelength from the unshocked SN ejecta arise from a planar region.

\begin{figure}[h]
\centering
\includegraphics[width=0.49\textwidth]{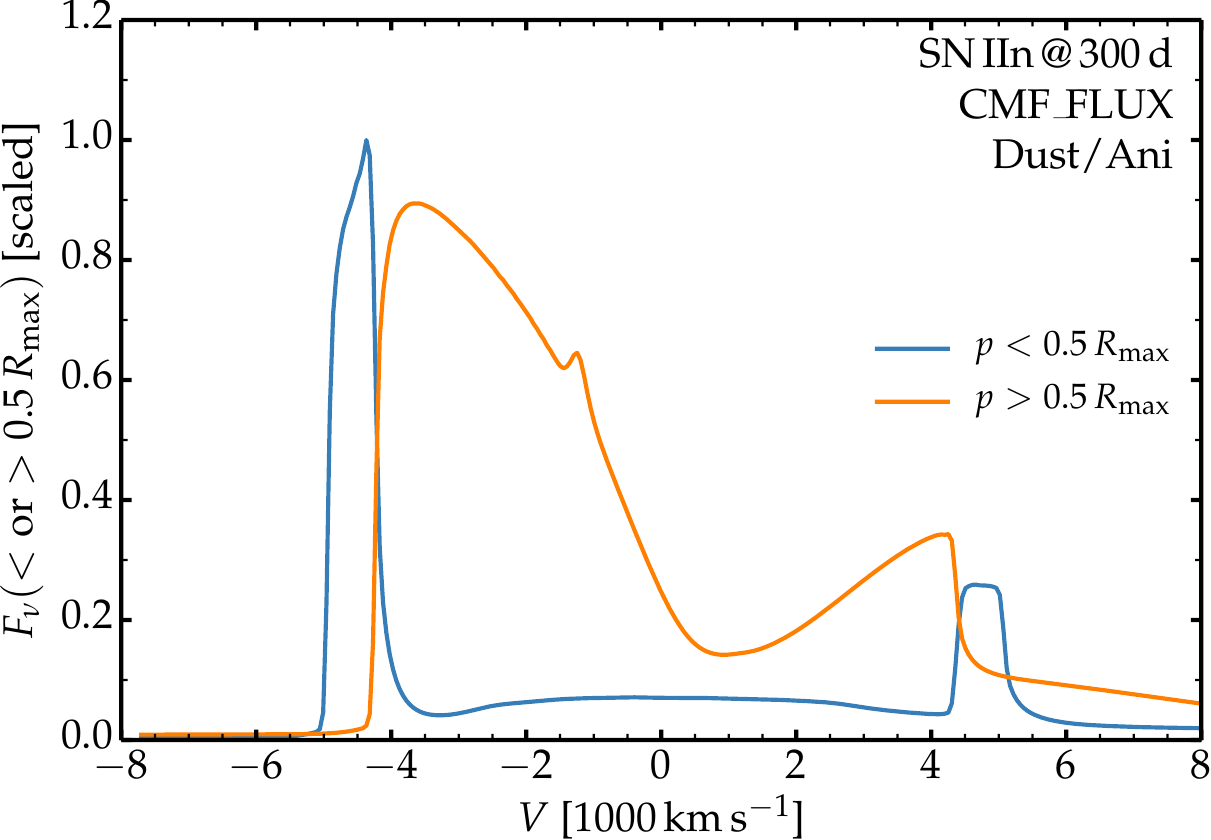}
\caption{Illustration for the dusty SN\,IIn model at 300\,d of the contribution to the escaping flux arising from ejecta regions with impact parameters smaller and greater than 0.5\,$R_{\rm max}$.
\label{fig_comp_mc_cmfgen_gp8_spec}
}
\end{figure}

 Figure~\ref{fig_comp_mc_cmfgen_gp8_spec} illustrates the H$\alpha$ profiles arising from regions with impact parameters smaller and greater than 0.5\,$R_{\rm max}$, where $R_{\rm max}$ is the radius of the dust shell. Both these latter figures are provided to illustrate the fundamentally different formation of SN ejecta line profiles, and those formed in a dense shell.

\section{Ejecta properties in the SN\,IIn model}

\begin{figure}[h]
\centering
\includegraphics[width=0.49\textwidth]{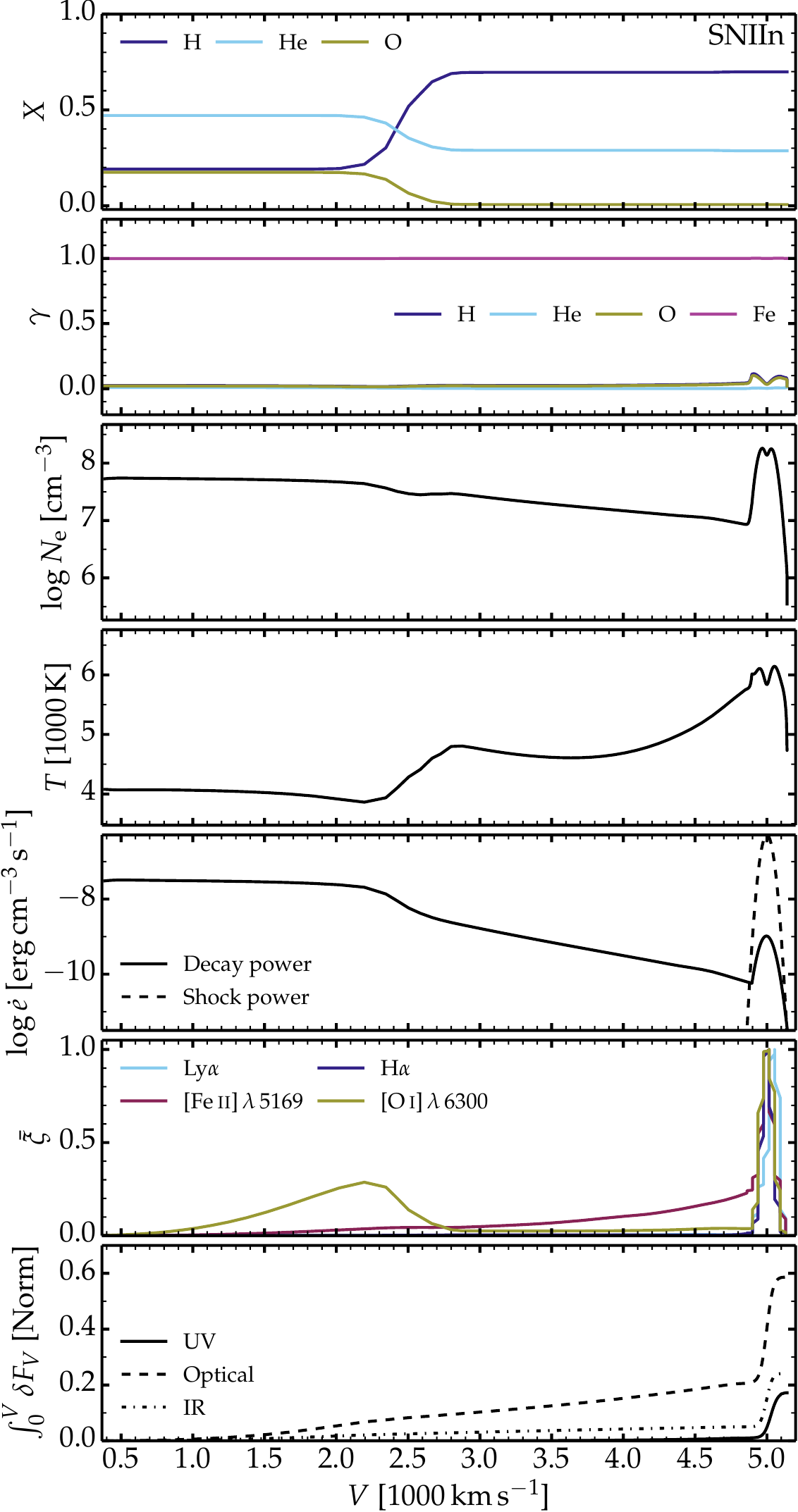}
\caption{Illustration of the ejecta and radiation properties versus velocity in the SN\,IIn model at 300\,d after explosion and used for comparison to SN\,1998S at 375\,d in Section~\ref{sect_sniin}. The main feature of this model is the presence of a massive dense shell at 5000\,\kms\ and powered by interaction (the interaction power absorbed is $2 \times 10^{41}$\,\ergs\ in this model). From top to bottom, we show the mass fractions and ionization state of H, He, O, and Fe (a value of zero corresponds to a neutral state, of one to once ionized etc), the free-electron density, the electron temperature, the absorbed powers from radioactive decay and ejecta/CSM interaction, some line emission measure denoted $\bar{\zeta}$ ($\zeta$ is defined such that $\int_a^b \zeta dr$ gives the fractional line flux originating between radii $a$ and $b$, and for better visibility we show its scaled value $\bar{\zeta}$; for details see \citealt{hillier_87}), and the outward integral of the flux in the ultraviolet, optical, and infrared. \label{fig_m15_300d_prop}
}
\end{figure}

We provide additional information about the ejecta properties of our SN\,IIn model in Fig.~\ref{fig_m15_300d_prop}. For the other two models, the ejecta properties are discussed in \citet[noninteracting SN\,II model]{dessart_sn2p_21} and in \citet[SN\,II model with late-time interaction with a standard red-supergiant wind]{dessart_late_23}.

\end{document}